\documentclass[10pt,journal,compsoc]{IEEEtran}
\usepackage{amssymb}
\usepackage[ruled,vlined]{algorithm2e}
\usepackage[justification=centering]{caption}
\usepackage{booktabs}
\usepackage{multirow}
\usepackage{url, color, xcolor}
\usepackage{amssymb}
\usepackage{bbding}
\usepackage{todonotes}
\usepackage{listings}
\usepackage{hyperref}

\ifCLASSOPTIONcompsoc
  \usepackage[nocompress]{cite}
\else
  \usepackage{cite}
\fi
   \usepackage{graphicx}
\usepackage{amsmath}
\usepackage{algorithmic}

\usepackage{array}

\usepackage{stfloats}

\usepackage{todonotes}

\begin{document}
%
\title{K-ST: A Formal Executable Semantics of the Structured Text Language for PLCs}
%
%
%
%

\author{Kun~Wang,
        Jingyi~Wang,
        Christopher~M.~Poskitt,
        Xiangxiang~Chen,
        Jun~Sun,
        and~Peng~Cheng
\IEEEcompsocitemizethanks{\IEEEcompsocthanksitem K. Wang, and J. Wang are with the College of Control Science and Engineering, Zhejiang University, Zhejiang 310027, China.\\
E-mail: \{kunwang\_yml, wangjyee\}@zju.edu.cn.
\IEEEcompsocthanksitem CM. Poskitt is with the School of Computing and Information Systems, Singapore Management University, Singapore.\\
E-mail: cposkitt@smu.edu.sg.
\IEEEcompsocthanksitem X. Chen is with the College of Control Science and Engineering, Zhejiang University, Zhejiang 310027, China.\\
E-mail: chenxiangx@zju.edu.cn.
\IEEEcompsocthanksitem J. Sun is with the School of Computing and Information Systems, Singapore Management University, Singapore.\\
E-mail: junsun@smu.edu.sg.
\IEEEcompsocthanksitem P. Cheng is with the College of Control Science and Engineering, Zhejiang University, Zhejiang 310027, China.\\
E-mail: lunarheart@zju.edu.cn.
}
\thanks{(Corresponding authors: Jingyi Wang and Peng Cheng)}
}

\IEEEtitleabstractindextext{
\begin{abstract}
Programmable Logic Controllers~(PLCs) are responsible for automating process control in many industrial systems (e.g.~in manufacturing and public infrastructure), and thus it is critical to ensure that they operate correctly and safely.
The majority of PLCs are programmed in languages such as Structured Text~(ST). However, a lack of formal semantics makes it difficult to ascertain the correctness of their translators and compilers, which vary from vendor-to-vendor.
In this work, we develop K-ST, a formal executable semantics for ST in the $\mathbb{K}$ framework.
Defined with respect to the IEC 61131-3 standard and PLC vendor manuals, K-ST is a high-level reference semantics that can be used to evaluate the correctness and consistency of different ST implementations. We validate K-ST by executing 567 ST programs extracted from GitHub and comparing the results against existing commercial compilers (i.e., CODESYS, CX-Programmer, and GX Works2).
We then apply K-ST to validate the implementation of the open source OpenPLC platform, comparing the executions of several test programs to uncover five bugs and nine functional defects in the compiler.
\end{abstract}

\begin{IEEEkeywords}
Formal executable semantics, PLC programming, Structured text, $\mathbb{K}$ framework, OpenPLC.
\end{IEEEkeywords}}

\maketitle

\IEEEdisplaynontitleabstractindextext

%
\IEEEpeerreviewmaketitle

\ifCLASSOPTIONcompsoc
\IEEEraisesectionheading{\section{Introduction}\label{sec:introduction}}
\else
\section{Introduction}
\label{sec:introduction}
\fi

%
%
%
%

\IEEEPARstart{P}{rogrammable} Logic Controllers~(PLCs) are responsible for automating process control in several modern industrial systems, e.g.~in manufacturing and public infrastructure.
It is critical to ensure that PLCs are operating correctly, as any functional or security-related defects may lead to serious incidents in the system.
This has most famously been demonstrated by the Stuxnet worm~\cite{langner2011stuxnet}, while many other less-known safety and security incidents~\cite{liang20162015,zetter2017ukrainian,perlroth2018cyberattack} and potential hazards~\cite{tychalas2018open,nochvay2019security} related to PLCs have resulted in significant consequences, with an estimated \$350,000 in damage on average~\cite{tychalas2021icsfuzz}.

The majority of PLCs are programmed using languages defined in the IEC 61131-3 open international standard~\cite{IECStandard}.
Programs can be written in graphical languages such as Function Block Diagrams~(FBD), but the standard also defines {Structured Text}~(ST), a fully textual language based on the idea of organizing code into `function blocks' and designed with a syntax similar to Pascal.
ST is a particularly important IEC 61131-3 language given its utility for data processing~\cite{antonsen2020plc}, and the fact that snippets of ST are actually required in FBD and other graphical languages.
It is therefore important that translators and compilers for ST are correctly implemented and exhibit only expected behaviors when the code is being run on a PLC.

\begin{figure}[!t]
    \centering
    \includegraphics[width=\linewidth]{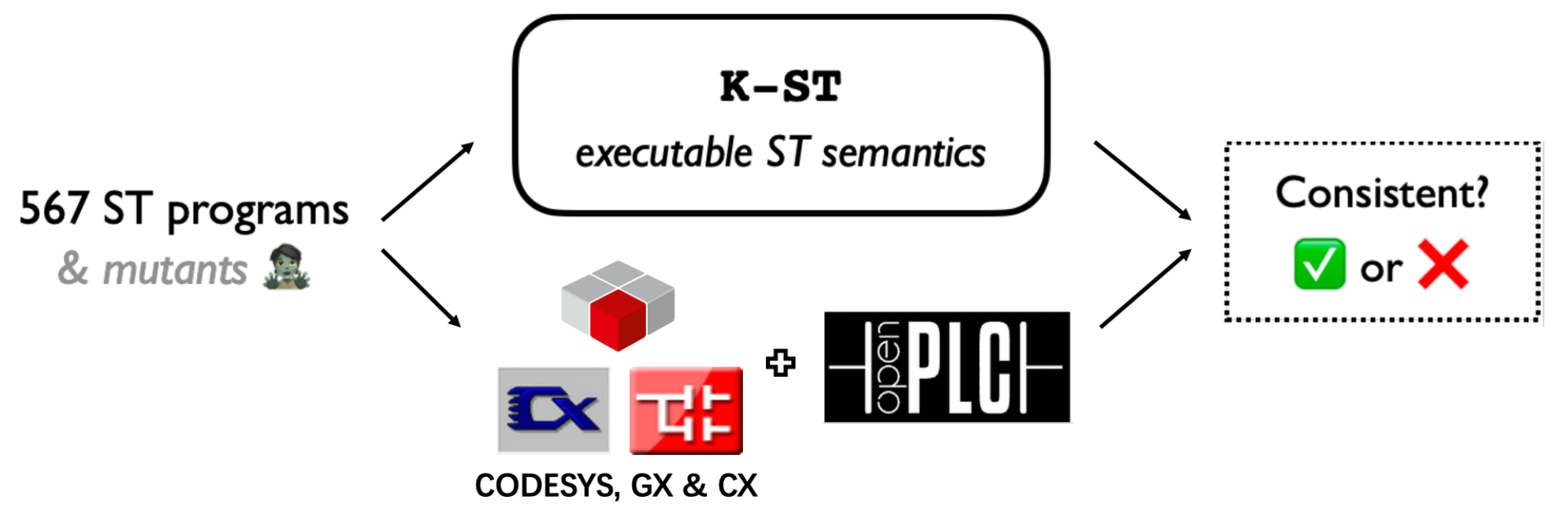}
    \caption{High-level workflow of our approach}
    \label{fig:workflow}
\end{figure}

This has motivated a surge of research on analyzing and verifying PLC programs \cite{blech2013formal,blech2011verification,tychalas2021icsfuzz,ovatman2016overview,janicke2015runtime,garcia2016detecting,darvas2014formal,darvas2016formal,garcia2017hey,spenneberg2016plc,keliris2018icsref,guo2017symbolic,mclaughlin2014trusted,canet2000towards,xiong2020safety,zhang2019towards,bauer2004verification,mader1999timed,mertke2001formal,huuck2005semantics,sadolewski2011conversion,adiego2014modelling}, although few works focus on ST implementations/compilers.
Zhang et al.~\cite{zhang2019towards} propose VetPLC, a temporal context-aware, program-based approach to produce timed event sequences that can be used for automatic safety vetting.
McLaughlin et al.~\cite{mclaughlin2014trusted} propose TSV which translates assembly-level code into an intermediate language (ILIL) to verify safety-critical code executed on PLCs.
Mader and Wupper~\cite{mader1999timed} translate Instruction List (IL) code into timed automata~\cite{maler1996hardware}. 
Bauer et al.~\cite{bauer2004verification} similarly use timed automata as the formalism for Sequential Function Chart (SFC). 
In~\cite{mertke2001formal}, the proposed method transforms IL to Petri-nets~\cite{heiner1997petri}, and manually builds two additional Petri-nets for modeling the PLC and its environment.
Xiong et al.~\cite{xiong2020safety} propose an algorithm based on variable state analysis for automatically extracting a {Behavior Model} (BM) from an ST program.
These works attempt to transform PLC programs into an intermediate language or another programming language (i.e., C) which is suitable for verifying or detecting potential issues using existing associated verifiers or checkers. The issue of these approaches is that they \emph{lack analysis and proof of equivalence in the conversion process.}
In addition, the analyses they perform are often limited (since the existing tools are not designed for PLCs) and do not offer the feedback to the level of source code.
Canet et al.~\cite{canet2000towards} propose formal semantics for a significant fragment of the IL language, and a direct coding of this semantics into a model checking tool.
Huuck~\cite{huuck2005semantics} develops a formal operational semantics and abstract semantics for IL, which allows approximating program simulation for a set of inputs in one simulation run.
Blech et al.~\cite{blech2011verification, blech2013formal, adiego2014modelling} attempted to define the formal semantics of the IL and SFC languages in Coq and NuSMV and, based on that, verify the safety properties in the code.
However, IL is a low level assembly-like language that has been deprecated from the IEC61131-3 standard. Furthermore, these studies mainly concentrate on analyzing the functional aspects of the programs and may overlook potential vulnerabilities and security risks introduced during the compilation process.

While extensive research has been conducted on testing more `traditional' compilers (e.g.~vulnerability detection for GCC and Clang~\cite{le2014compiler,yang2011finding,chen2020survey}), compilers for PLC languages such as ST have received much less attention.
The challenges associated with testing the implementation of a compiler arise from the inherent difficulties of ensuring its correctness. One particular challenge stems from the absence of a precise specification of the expected behavior of a compiler. For most popular programming languages, there exist multiple purportedly equivalent implementations of compilers. Compiler testing can take advantage of this by utilizing these implementations as oracles for conducting differential testing~\cite{mckeeman1998differential}. However, in the case of the domain-specific ST language, there is no specific implementation standard, and different vendors often develop their own compilers based on their specific requirements. Another challenge is the semantic complexity of the input and output languages that compilers handle. The fact that different vendors develop their own implementations further exacerbates this issue. Compiler testing methods based on formal semantics~\cite{schumi2021spectest} have shown advantages in addressing these challenges.
With a formal semantics of the ST language, the expected behavior of ST compilers can be precisely and unambiguously defined, which can greatly aid in testing and verifying their correctness.

To the best of our knowledge, a practical and complete semantics for the ST language does not exist, which makes it difficult to ascertain the correctness of ST translators and compilers (e.g.~by comparing executions).
There are a number of reasons why such a reference semantics is yet to emerge.
First, there is insufficient documentation defining or describing the complete features of the ST language~\cite{antonsen2020plc}.
For instance, the official documentation introduces language features by only a few examples, based on which it is difficult for readers to fully understand the behavior of the language.
Second, the ST compilers provided by different vendors (e.g.~Allen-Bradley, Siemens) can implement the language differently, and their closed source solutions make it difficult to fully assess how they behave systematically (other than through  manual observation).
{For example, CODESYS, CX-Programmer, and GX Works2 all produce negative numbers in the results of negative modulo operations, even though this behavior is undefined according to the official documentation. Furthermore, GX Works2 supports only 10 basic data types, whereas CODESYS supports 17 types. Thus, a formal semantics needs to be `concrete' enough to be useful, but `high-level' enough to be general/extendable to the different nuances of vendors’ compilers.}
A preliminary attempt at defining a high-level semantics for ST was made by Huang et al.~\cite{huang2019kst}. However, it falls short of a full reference semantics as it misses several important features of the language, e.g.~certain data types, and key sentences.

In this work, we develop K-ST, a formal executable reference semantics for ST in the $\mathbb{K}$ framework~\cite{rosu2017k}.
Our high-level semantics is both executable and machine readable, and can be used by the $\mathbb{K}$ framework to generate interpreters, compilers, state-space explorers, model checkers, and deductive program verifiers.
Our principal goals for the design of K-ST are as follows:

\begin{enumerate}
    \item \textbf{Validated reference semantics.} K-ST is designed to cover all the main features of ST, and is validated against hundreds of different real-world ST programs extracted from GitHub.
    \item \textbf{General and extendable.} The semantics is high-level (rather than tied to a particular compiler), with the goal of supporting different ST implementations as well as extensions for vendor-specific functions.
    \item \textbf{Analyses of ST compilers.} Most importantly, K-ST can be used to check the correctness and consistency of different ST implementations, and thus ensure that a compiler is not introducing an unintended behavior or compile-time threat~\cite{hohnka2019evaluation,marcozzi2019compiler} into a critical industrial system.
\end{enumerate}

Given the absence of complete feature descriptions for the ST language in official documentation, we not only refer to the definitions and code samples in the official documents, but also extensively consult the guidance manuals provided by multiple vendors to better define the semantics of the ST language.
For example, there is no specific documentation on how integer overflow is handled in the official documents. Through investigating multiple instruction manuals, we found that existing ST compilers generally use truncation to handle integer overflow without any warning. In defining the semantics, we find that the rewriting rule of the $\mathbb{K}$ framework provides a good mechanism for capturing the unique features of ST. For example, we can rewrite $\mathtt{REPEAT}$ to $\mathtt{WHILE}$ to achieve the execution effect of $\mathtt{REPEAT}$.

We validate K-ST by extracting 567 real-world ST code samples from GitHub and comparing their executions in our semantics against their executions resulting from various commercial compilers (i.e., CODESYS, CX-Programmer, and GX Works2). We find that K-ST is sufficiently complete to support 509 of these programs (consisting of 26,137 lines of code) and executes those programs correctly (i.e., producing the same outputs as the corresponding existing compiler), with the remaining programs only unsupported due to the use of certain vendor-specific or hardware-related functions that we did not yet formalize.
Furthermore, to evaluate the utility of K-ST for testing ST compilers, we compared the executions of the 567 programs (and several mutants) under K-ST and OpenPLC~\cite{alves2014openplc}, a popular open source PLC program compiler.
Through this semantics-based testing, we are able to uncover five bugs and nine functional defects in the OpenPLC compiler, all of them are previously unknown.
Fig.~\ref{fig:workflow} summarises the high-level workflow of this process.

In summary, we make three main contributions.
\begin{itemize}
    \item We propose an executable formal reference semantics for ST;
    \item We collect a set of 567 complete ST program samples, and validate the correctness of our executable semantics by running those programs in the semantics and via existing compilers (CODESYS, CX-Programmer, and GX Works2), comparing the results. 
    \item We test OpenPLC, an open source PLC program compiler, using our proposed semantics, and find five bugs and nine functional defects.
\end{itemize}

The remaining part of this paper is organized as follows. Section~\ref{sec:background} introduces the background of ST and the $\mathbb{K}$ framework. The proposed executable operational semantics of ST formalized in $\mathbb{K}$ is introduced in Section~\ref{sec:kst}. Section~\ref{sec:applications} shows some practical applications of our formal semantics. The evaluation results of the proposed semantics are introduced in Section~\ref{sec:evaluation}. Section~\ref{sec:related_work} summarises some related work, and Section~\ref{sec:conclusion} concludes this work.


\section{Background}
\label{sec:background}
In this section, we briefly introduce the background of the Structured Text~(ST) language and the $\mathbb{K}$ framework.

\subsection{Structured Text}
The Programmable Logic Controller, invented in 1969 by Dick Morley, is specially designed for applications in industrial environments, e.g. assembly lines, robotic devices, or public infrastructure. 
These kinds of applications all require high reliability and ease of programming.

\begin{figure}[!t]
\centering
\includegraphics[width=0.49\textwidth]{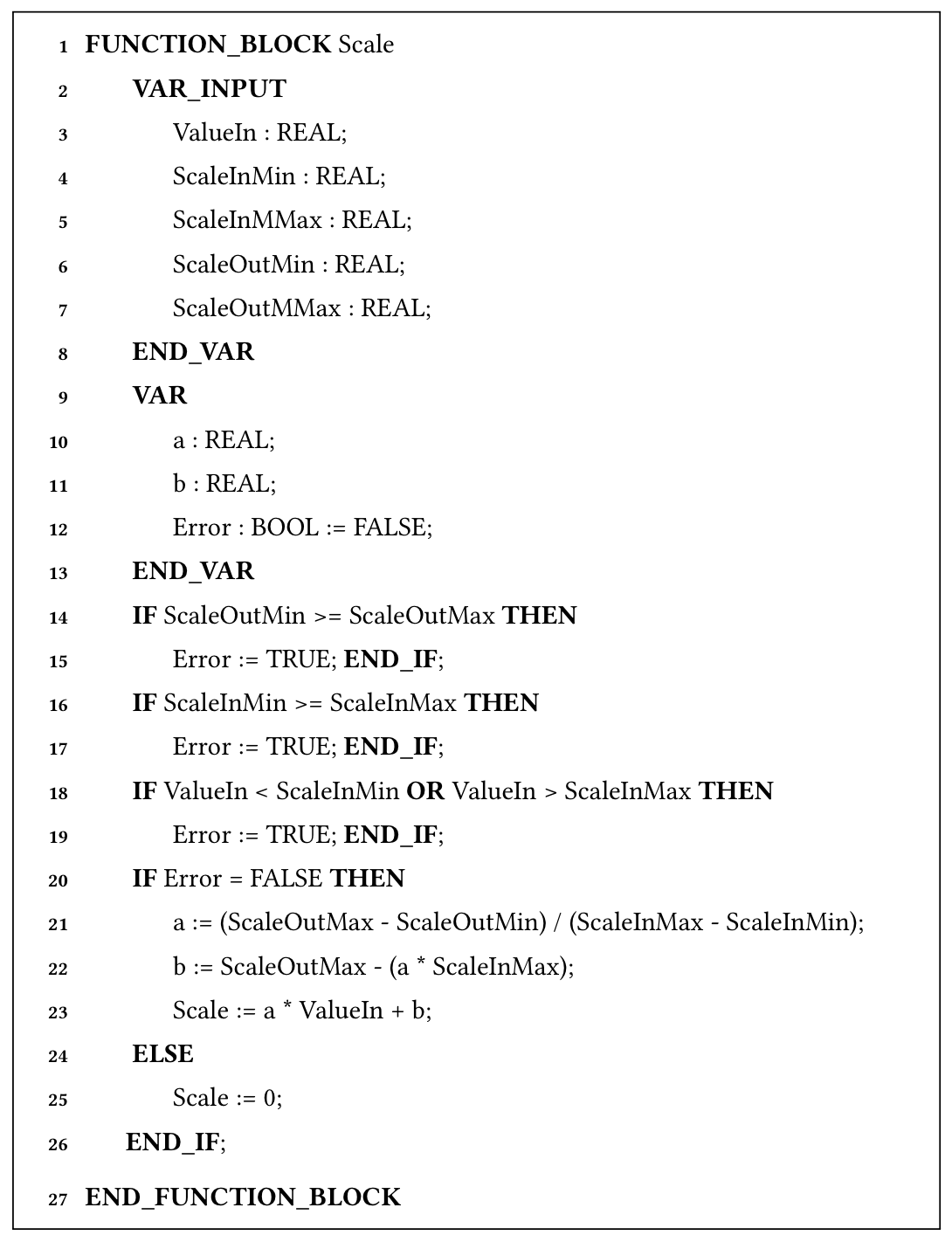}
\caption{An ST programming example} \label{fig1}
\end{figure}

Early PLCs were represented as a series of logic expressions in some kind of Boolean format. With the development of programming terminals and the complexity of existing control procedures, Ladder Diagrams (LD) were developed to program PLCs. As of 1993, the IEC 61131-3 standard developed by the International Electrotechnical Commission (IEC) defined five programming languages, including two textual programming languages---ST and IL---as well as three graphical languages---LD, FBD, and SFC. A simple example in Fig.~\ref{fig1} \cite{darvas2016generic} shows a ST code example which can be used for linear scaling of an analog sensor signal.

ST is a high-level PLC programming language which is similar to Pascal \cite{roos2008programming} (widely used from 1980 to 2000), C/C++ and Java. While it contains common constructs from modern programming languages such as $\mathtt{FUNCTION}$, $\mathtt{IF/ELSIF/ELSE}$ and $\mathtt{CASE}$ branches, $\mathtt{WHILE}$ and $\mathtt{FOR}$ loops, it has its own characteristics, such as the lack of recursion, capitalized keywords, $\mathtt{REPEAT}$ statement, and $\mathtt{FUNCTION\_BLOCK}$ structure. 
For instance, $\mathtt{FUNCTION\_BLOCK}$ as an important part of ST,  and has its own state. Its main purpose is to modularize and structure a straightforwardly defined portion of the program. It is similar to the class-object manifestation in {object-oriented} programming. Function blocks exist in two forms: as a type or as an instance, but only the instance can be called. For each function block, the local variables retain their values between each `call'. TABLE~\ref{tab:element} shows the common elements of ST.
\begin{table*}[!t]
\centering
  \caption{Common elements of the ST language}
  \label{tab:element}
  \renewcommand\arraystretch{1.5}
  \begin{tabular}{cc|cc|cc}
    \toprule
    Type & Element & Type & Element & Type & Element\\
    \midrule
    \multirow{3}{*}{Program Organization Unit} & FUNCTION\_BLOCK & \multirow{15}{*}{Built-in Data Type} & INT & \multirow{2}{*}{Built-in Data Type} & ARRAY\\
                              & FUNCTION &   & DINT &                 & ... \\
    \cline{5-6}
	                          & PROGRAM &           & SINT &  \multirow{12}{*}{Declaration Type} & VAR\_GLOBAL\\
	\cline{1-2}
    \multirow{7}{*}{Main Statement} & IF & & LINT & & VAR\\
    & CASE & & UINT & & VAR\_INPUT\\
    & WHILE & & UDINT & & VAR\_OUTPUT\\
    & FOR & & USINT & & VAR\_IN\_OUT\\
    & REPEAT & & ULINT & & VAR\_EXTERNAL\\
    & EXIT & & REAL & & VAR\_TEMP\\
    & RETURN & & LREAL & & AT\\
    & ... & & BOOL & & RETAIN\\
    \cline{1-2}
    \multirow{2}{*}{User Data Type} & ENUM & & STRING & & PERSISTENT\\
    & STRUCT & & WSTRING & & CONSTANT\\
    \cline{1-2}
    \multirow{2}{*}{Built-in Data Type} & TIME & & TIME\_OF\_DAY & & ...\\
    \cline{5-6}
    & DATE & & DATE\_AND\_TIME & \multicolumn{2}{c}{...}\\
  \bottomrule
\end{tabular}
\end{table*}

ST, as the only textual programming language supported by the new IEC standard, has a number of advantages compared to other PLC languages. First, being textual, ST programs can be copied relatively easily. Second, compared with the other four languages, it is more convenient for mathematical calculations, formulas and algorithms, and for managing large amounts of data \cite{antonsen2020plc}. Third, compared with 20 years ago, PLC solutions are more in demand today and ST can better adapt to this change. Finally, LD, SFC and FBD also require parts of the program to be written in ST anyway \cite{markovic2015automated,tiegelkamp2010iec}. 

Unfortunately, the absence of documents defining or describing the complete features of the ST language and the differing customizations of vendors can lead to inconsistent implementations of ST. In addition, understanding the semantics of the ST language, and ensuring that it is formally defined is difficult for end users accustomed to graphical programming. A formal executable semantics of ST not only provides a standard, but also helps PLC engineers verify the completeness and correctness of these implementations.

\subsection{The $\mathbb{K}$ Framework}
$\mathbb{K}$ is a formal logic framework based on rewriting logic \cite{marti2002rewriting}. It was developed with the overarching goal of pursuing the ideal language framework, where all programming languages have formal semantic definitions and all language tools are automatically derived in a correct-by-construction manner at no additional cost. The $\mathbb{K}$ backends, such as the Isabelle theory generator, the model checker, and the deductive verifier, can be utilized to prove properties based on the semantics and generated verification tools \cite{stefuanescu2016semantics}. Several executable semantics in $\mathbb{K}$ have been developed for mainstream programming languages, including C~\cite{ellison2012executable}, Java~\cite{bogdanas2015k}, JavaScript~\cite{park2015kjs}, Rust~\cite{wang2018krust}, Solidity~\cite{jiao2020semantic}, and IMP~\cite{nipkow2014imp}.

A language semantics definition in $\mathbb{K}$ consists of three parts: the language syntax, the configuration, and a set of semantics constructed based on the syntax and the configuration. Given the semantics definition for a programming language and some source programs, $\mathbb{K}$ executes these programs like a translator. For illustration, in the following we take a strict subset of the ST language, i.e., ST$_{demo}$ shown in Fig.~\ref{fig1} as an example to illustrate how to define language semantics in $\mathbb{K}$. 

\textbf{\textit{Configuration.}} The whole configuration cell $T$ of ST$_{demo}$ contains two cells, namely $k$ and $state$. The cell $k$ is used to store the source program $\$PGM$ for execution, and the cell $state$ is used to record the mapping from a variable identifier to its value. The configuration simulates the memory status and environmental changes during runs of the program.
$$\left\langle \left\langle \$PGM:Pgm \right\rangle _k \; \left\langle .Map \right\rangle _{state} \right\rangle _T$$

\begin{figure}[!t]
\centering
\includegraphics[width=0.48\textwidth]{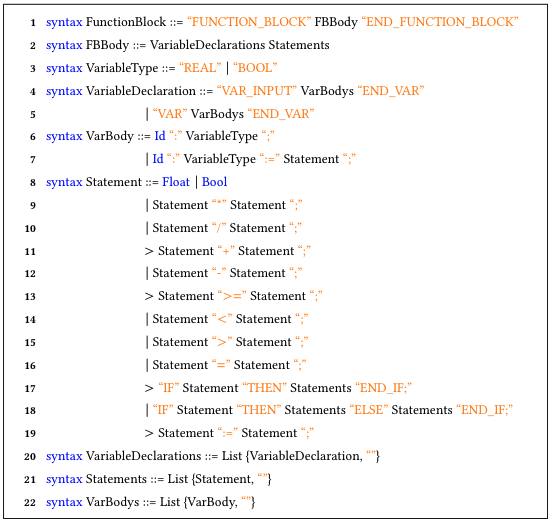}
\caption{The syntax of ST$_{demo}$} \label{fig2}
\end{figure}

With the configuration defined, we present the syntax of ST$_{demo}$ in Fig.~\ref{fig2}, which includes some numerical operations, logic operations and commonly used statements. Based on the configuration and the syntax of ST$_{demo}$, we introduce some basic rules in the semantics. The role of the semantics is to tell $\mathbb{K}$ how to execute the source code, where $\mathbb{K}$ executes the code and updates the configuration sentence-by-sentence after parsing the source program.

\begin{figure}[!t]
\centering
\includegraphics[width=0.48\textwidth]{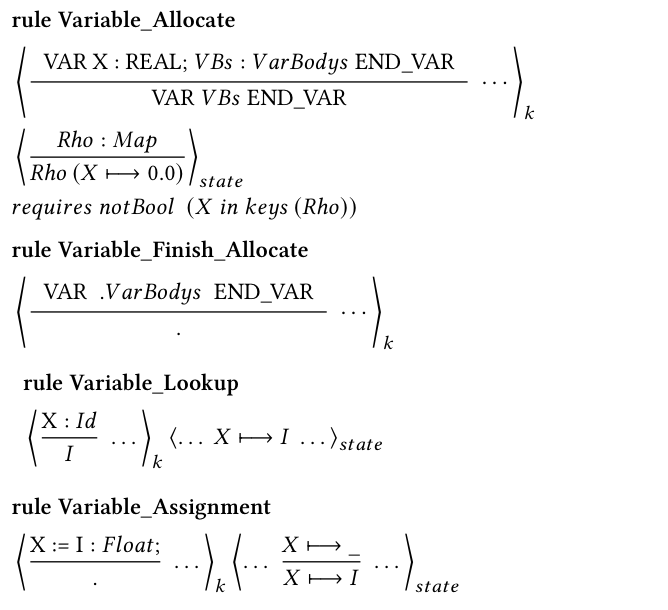}
\caption{The partial semantics of ST$_{demo}$} \label{fig9}
\end{figure}

Here, we show the semantics of $Allocate$, $Lookup$ and $Assignment$ in Fig.~\ref{fig9} as they are the most commonly used constructs in programming languages. TABLE~\ref{tab:fuhao} describes some common semantic notations.
Take $Allocate$ as an example: when $\mathbb{K}$ runs to lines 9--13 in Fig.~\ref{fig1}, the content in the $k$ cell is $\left \langle \mathtt{VAR\;a:REAL;\;}VBs\mathtt{\;END\_VAR \;} \cdots \right\rangle _k $, where $VBs$ stands for $\mathtt{b:REAL;\;Error:BOOL:=FALSE;}$. Then, $\mathbb{K}$ will rewrite $\left \langle \mathtt{VAR\;a:REAL;\;}VBs\mathtt{\;END\_VAR \;} \cdots \right\rangle _k $ to $\left \langle \mathtt{VAR\;}VBs\mathtt{\;END\_VAR \;} \cdots \right\rangle _k $, which means that $\mathtt{a:REAL;}$ has been executed according to $\mathbf{rule \; Variable\_Allocate}$. Meanwhile, it adds the mapping between the variable name and the corresponding value ($a \mapsto 0.0$) in the current $state$ cell $Rho$. In addition, ``$requires \; notBool \; \left ( X \; in \; keys\left ( Rho \right )  \right )$'' guarantees that the variable will not be re-declared. Similarly, variables $b$ and $Error$ will be allocated separately. After that, the content in the $k$ cell is $\left \langle \mathtt{VAR\;}.VarBodys\mathtt{\;END\_VAR \;} \cdots \right\rangle _k $, where $.VarBodys$ represents an empty variable declaration list, that is, no additional variable needs to be allocated. The $\mathbf{rule\;Variable\_Finish\_Allocate}$ will be called to convert ``$\text{VAR}\;.VarBodys\;\text{END\_VAR}$'' in $k$ to ``$.$'', which means that there is no more code to execute in the $\text{VAR}$ block and $\mathbb{K}$ will continue to execute the subsequent code.

\begin{table}[!t]
    \centering
    \caption{Summary of semantic notations}
    \renewcommand\arraystretch{1.5}
    \begin{tabular}{c|l}
        \toprule
        Notation & \multicolumn{1}{c}{Description}\\
        \midrule 
        $\mathbf{rule}$ & The beginning of a semantic rule.\\
        \hline
        $a \Rightarrow b$ & \begin{tabular}[c]{@{}l@{}}The symbol $\Rightarrow$ means ``rewritten by'', thus $a \Rightarrow b$\\denotes that $a$ can be replaced by $b$.\end{tabular}\\
        \hline
        $a \; requires \; b$ & Execute $a$ when $b$ is true.\\
        \hline
        \multirow{3}{*}{$ \left \langle \frac{a}{b} \right \rangle _k$} & $\left \langle \right \rangle_k$ stands for the $k$ cell in a configuration.\\
        \cline{2-2}
        & Similar to $a \Rightarrow b$, $\frac{a}{b}$ means $a$ will be rewritten by $b$. \\
        &However, it can only be used inside $\left \langle \right\rangle$.\\
        \hline
        $\left \langle \cdots \; a \; \cdots \right\rangle$ & $\cdots$ represents the content in the $a$ context.\\
        \hline
        $.$ & $.$ stands for empty.\\
        \hline
        $\_$ & Any value.\\
        \hline
        $a \; : \; b$ & The type of variable $a$ is $b$.\\
        \hline
        $a \mapsto b$, $a \gets b$ & Mapping from $a$ to $b$.\\
        \hline
        $a \; \curvearrowright \; b$ & The execution of $a$, followed by execution of $b$.\\
    \bottomrule
    \end{tabular}
    \label{tab:fuhao}
\end{table}

\section{Formal Semantics of Structured Text in The $\mathbb{K}$ framework}
\label{sec:kst}
In this section, we introduce K-ST, the executable operational semantics of ST formalized in $\mathbb{K}$. Note that in practice the PLC programming environment is provided by specific PLC manufacturers including {CODESYS} and Siemens's TIA portal (TIA, Structured Control Language (SCL)). As a consequence, the implementations of different manufacturers can vary and may also include their own unique functions or structures. 

Our approach is therefore to focus on the common features, allowing other unique functions of the environment to be implemented by extending the 
operational semantics. Specifically, the syntax of ST is constructed based on the official IEC 61131-3 standard \cite{tiegelkamp2010iec}. The configuration is specifically designed for ST. Based on the syntax and the configuration, we then formalize the semantic rules for the language features with rewriting logic. Next, we present each component of the semantic one by one.

\subsection{The Syntax of ST}
TABLE~\ref{tab:syntax} presents the syntax of ST defined in K-ST, which covers most of the core syntax. We remark that
TABLE~\ref{tab:syntax} only contains the main part of K-ST while omitting others, e.g., some built-in functions ($\mathtt{LEN}$, $\mathtt{DELETE}$ and so on) for space reasons.
The syntax is specified by a grammar in a dialect of Extended Backus-Naur Form (EBNF)~\cite{mccracken2003backus}, where $^*$ means zero or more repetitions. In ST, the top-level grammatical structures include user-defined types ($\mathtt{TYPE}$ statements) and three Program Organization Units (POUs): $\mathtt{FUNCTION}$, $\mathtt{FUNCTION\_BLOCK}$ and $\mathtt{PROGRAM}$. 
Other syntactical elements are derived within these top-level grammatical structures.

\begin{table*}[!t]
\newcommand{\tabincell}[2]{\begin{tabular}{@{}#1@{}}#2\end{tabular}}
	\caption{The syntax of ST}
    \label{tab:syntax}
    \renewcommand\arraystretch{1.5}
    \begin{tabular}{ll}
    \toprule
    Syntax & Description\\
    \midrule
    \tabincell{l}{$Id ::= \left [ a-zA-z\_ \right ] \left [ a-zA-Z0-9\_ \right ]^*$ \\ 
         $Ids ::= Id^*$\\
         $IdVal ::= Id := Expression$} & Identifier \\
    \hline
    \tabincell{l}{$EnumStructDeclaration ::= TYPE \; EnumDeclarationExp^* \; END\_TYPE$\\
        $| \; TYPE \; StructDeclarationExp^* \; END\_TYPE$\\
        $EnumBlock ::= Ids \; | \; IdVal^* $\\
        $EnumDeclarationExp ::= Id : \left( EnumBlock \right); \; | \; Id : \left( EnumBlock \right) := Id;$\\
        $StructDeclarationExp ::= Id : STRUCT \; VarDeclarationExp^* \; END\_STRUCT$} & Enum and Struct declaration\\
    \hline
    \tabincell{l}{$Function ::= FUNCTION \; Id \; : \; Type \; VarDeclaration^* \; Statements \; END\_FUNCTION$} & Function declaration\\
    \hline
    \tabincell{l}{$FunctionBlock ::= FUNCTION\_BLOCK \; Id \; VarDeclaration^* \; Statements \; END\_FUNCTION$} & Function block declaration\\
    \hline
    \tabincell{l}{$Program ::= PROGRAM \; Id \; VarDeclaration^* \; Statements \; END\_PROGRAM$} & Program declaration\\
    \hline
    \tabincell{l}{$Type ::= INT | DINT | SINT | LINT | UINT | UDINT | USINT | ULINT | BYTE | WORD | DWORD | REAL $ \\ $ | LREAL | STRING | STRING\left [ Expression \right ] | WSTRING | WSTRING \left [ Expression \right ]  | TIME | DATE $ \\ $| TIME\_OF\_DAY | DATE\_AND\_TIME | Id | ARRAY \left [ Expression \right ] \; OF \; Type$} & Variable types \\
    \hline
    \tabincell{l}{ $VarType ::= VAR\_GLOBAL \; | \; VAR \; | \; VAR\_INPUT \; | \; VAR\_OUTPUT \; | \; VAR\_IN\_OUT \; | \; VAR\_TEMP$ \\ $VarDeclarationExp ::= Ids : Type; \; | \; Ids : Type := Expression; $ \\
        $VarDeclaration ::= VarType \;\; VarDeclaration \;\; END\_VAR$} & Variable declaration \\
    \hline
	\tabincell{l}{$Operation ::= + \; | \; - \; | \; * \; | \; / \; | \; ** \; | \; MOD \; | \; < \; | \; > \; | \; = \; | \; <= \; | \; >= \; | \; <> \; | \; AND \; | \; \& \; $ \\
	    $| \; AND\_THEN \; | \; XOR \; | \; OR \; | \; OR\_ELSE \; | \; ..$ \\
	    $Expression ::= Int \; | \; Float \; | \; String \; | \; Bool \; | \; Bit \; | \; AllTime \; | \; Id \; | \; Expression \; Operation \; Expression $ \\
	    $Expression \left( Expressions \right) \; | \; Expression . Expression \; | \; Expression \left [ Expressions \right ] \; | \; \left( Expression \right)$\\
	    $Expressions ::= Expression^*$} & Expressions \\
	\hline
	\tabincell{l}{$Assignment ::= Expression := Expression;$} & Assignment statement\\ 
	\hline
	\tabincell{l}{$ElseIfBlock ::= ELSE \; Statements \; | \; ELSE\_IF \; Expression \; THEN \; Statements \; ElseIfBlock^* $\\
	$If ::= IF \; Expression \; THEN \; Statements \; ElseIfBlock^* \; END\_IF;$\\
	$CaseBlock ::= Expression  :  Statements \; | \; Expression \; .. \; Expression : Statements$\\
	$Case ::= CASE \; Expression \; OF \; CaseBlock^* \; END\_CASE; \; $\\
	$| \; CASE \; Expression \; OF \; CaseBlock^* \; ELSE \; Statements \; END\_CASE;$} & Branch statements \\
	\hline
	\tabincell{l}{$While ::= WHILE \; Expression \; DO \; Statements \; END\_WHILE;$\\
	$For ::= FOR \; Expression \; TO \; Expression \; DO \; Statements \; END\_FOR;$\\
	$| \; FOR \; Expression \; TO \; Expression \; BY \; Expression \; DO \; Statements \; END\_FOR;$\\
	$Repeat ::= REPEAT \; Statements \; UNTIL \; Expression \; END\_REPEAT;$} & Loop statements \\
	\hline
	\tabincell{l}{$Return ::= RETURN;$} & Return statement \\
	\hline
	\tabincell{l}{$Exit ::= EXIT;$} & Exit statement \\
	\hline
	\tabincell{l}{$Statement ::= Expression; \; | \; Assignment \;| \; If \; | \; Case \; | \; While \; | \; For \; | \; Repeat \; | \; Return \; | \; Exit $\\
	    $Statements ::= Statement^*$} & Statements \\
    \bottomrule
\end{tabular}
\end{table*}

\subsection{The Configuration of ST}
\begin{figure*}[!t]
\centering
\includegraphics[width=0.9\textwidth]{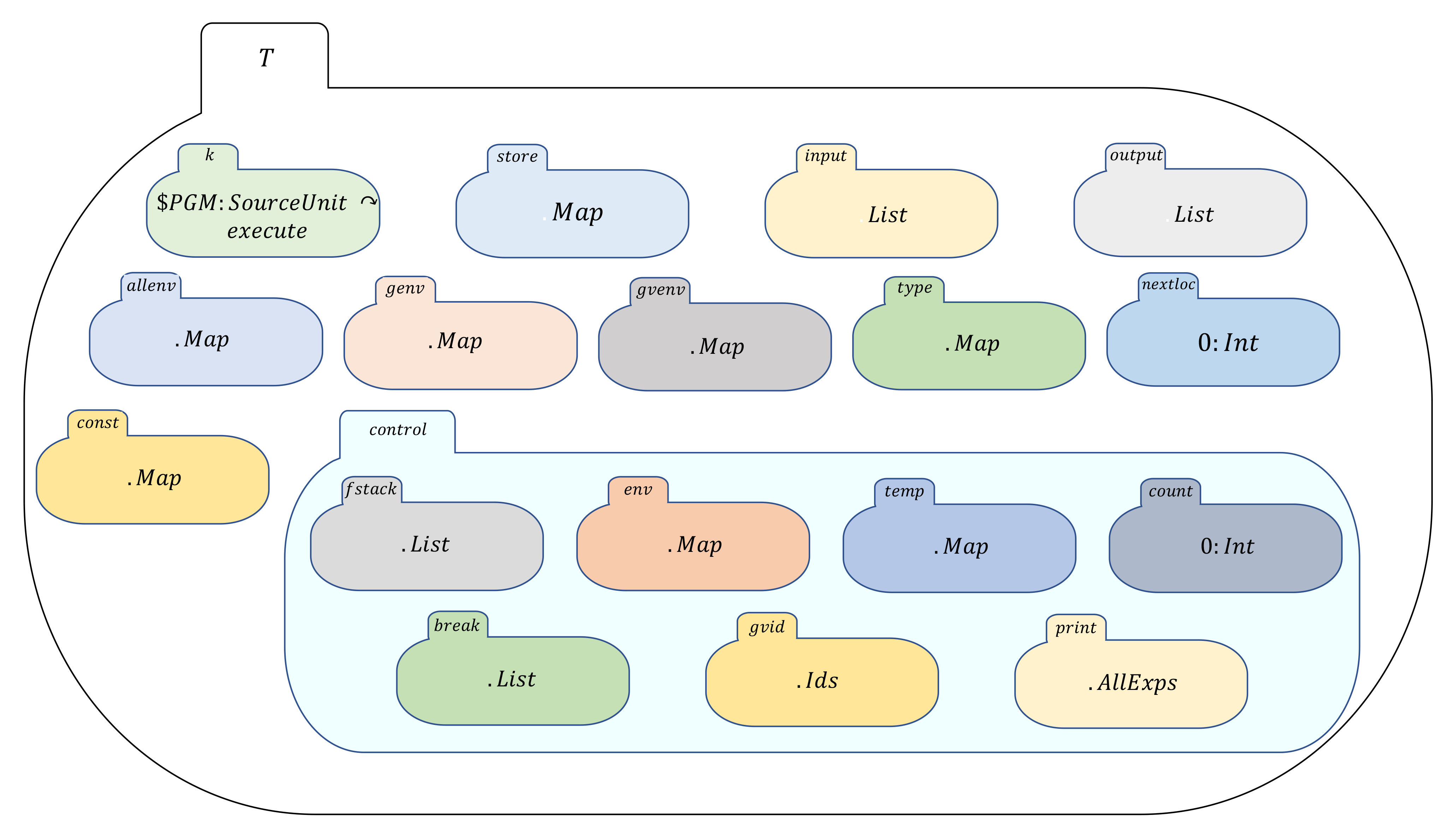}
\caption{The runtime configuration of ST in $\mathbb{K}$} \label{fig4}
\end{figure*}

The execution of an ST program needs to update the following kinds of state: data segment, code segment and stack. Among them, the data segment is used to store global variables, the code segment is used to store program execution code, and the stack is used to store local variables of the program. Note that runtime environment switching caused by function calls is also achieved by the operation of stack.
The overall runtime configuration of ST in $\mathbb{K}$ is presented in Fig.~\ref{fig4}. We highlight our careful
design choices as follows. 

    \textbf{\textit{Overview.}} There are 11 main cells in the configuration $T$, i.e., $k$, $control$, $allenv$, $genv$, $gvenv$, $store$, $type$, $constant$ $input$, $output$ and $nextLoc$. The value of each cell is initialized according to its specified type. For instance, for cells with a mapping relationship, their values are initialized to $Map$ type, and for cells that store a collection, they are initialized to $List$ type. A `$.$' followed by any type means an empty set of this type. For instance, $.Map$ in the cell $genv$ represents that $genv$ is initialized with an empty map.
    
    \textbf{\textit{Enumeration type.}} By default, when an enumeration type is defined in ST, PLC compilers will automatically associate a number (indexed from 0 and incremented by 1 each time) to each variable in the enumeration. For repeated declarations, we use $count$ cell to record the value of the current enumeration.  
    
    \textbf{\textit{Global variables.}} There are two types of global variables. First, the POUs and customized types that users define. These variables can be accessed anywhere in the program. We store these variables in the $genv$ cell as the basis for program operation. Second, the variables defined in $\mathtt{VAR\_GLOBAL}$. These variables cannot be directly accessed in the program unless they are declared with $\mathtt{VAR\_EXTERNAL}$. We store these variables in the $gvenv$ cell and provide them on demand. 
    
    \textbf{\textit{Program execution.}} The source code parsed by the syntax $SourceUnit$, called $\$PGM$, is stored in the cell $k$ for execution. Then the $\$PGM$ will be executed unit by unit. If the program terminates normally, there will be a `$.$' in the $k$ cell, denoting that no more units need to be executed. 
    In the preprocessing phase (the first pass of $\mathbb{K}$), the $k$ cell only contains the token $execute$. Afterwards, $\mathbb{K}$ will start executing from the $\mathtt{MAIN}$ program. 

    \textbf{\textit{Stack operations.}} The cell $control$ contains seven subcells---$fstack$, $env$, $temp$, $count$, $gvid$, $print$ and $break$---which record the operating environment of the currently running code segment. Specifically, the function stack $fstack$ is a list used to store the environment before executing other POUs, including variables in the current environment and the subsequent program. Next, the cell $env$ is used to store the mapping relationship between variables and indexes in the current environment during program execution. Furthermore, cells $temp$ and $count$ are used in $\mathtt{ENUM}$ and $\mathtt{STRUCT}$, where $temp$ is for temporary mapping and $count$ is used as a counting pointer. The cell $gvid$ records all identifiers of global variables to assist in the generation of global variables. The cell $print$ records variables which need to be output. Finally, $break$ stores the program after the loop in order to support the implementation of the $\mathtt{EXIT}$ statement in $\mathtt{FOR}$, $\mathtt{WHILE}$ and $\mathtt{REPEAT}$ loops.
    
    \textbf{\textit{Execution environment.}} The $allenv$ cell is used to cache the execution environment before function calls (for strict type checking of parameter passing in function calls\footnote{This is optional but recommended for ST compilers.}).
    The cell $genv$ records the result of the pre-processing (including POUs and custom types) and will be copied to $env$ when $env$ is refreshed. The last cell related to the environment is called $gvenv$ and is used to index global variables. 
    
    \textbf{\textit{Memory operation.}} The $store$ cell is used to simulate memory to record the mapping relationships of indexes and variable values. After that, the cells $input$ and $output$ are used to realize external inputs and external output respectively. The last cell, $nextLoc$, ensures that the index of a variable can always be incremented without duplication. The design consideration behind this is that for complex languages, it is more effective to explicitly manage arbitrarily large memory than use garbage collection \cite{rocsu2014k}.

\subsection{Semantics of the Core Features}
We implement the executable semantics covering most core features of ST and leave the vendor-specific functionalities as potential extensions. 
For example, some compilers would use additional keywords to distinguish the declaration part and the execution part of the program.
In the following, we provide an overview of four core semantic features of ST, including 1) data types, 2) main control statements, 3) declarations and calls of POUs and 4)  memory operations. 
Before diving into the details, we present the notations as follows.
\subsubsection{Extended Data Types}
The $\mathbb{K}$ framework supports diverse data types including identifiers ($Id$), integers ($Int$), bools ($Bool$), floats ($Float$) and strings ($String$), which cover most of the requirements.
However, there are still some unsupported data types needing additional implementation in K-ST, which we call extended data types. These extended data types can be categorized into two kinds: 1) elementary types ($\mathtt{TIME}$, $\mathtt{BYTE}$, $\mathtt{WORD}$, $\mathtt{DWORD}$, $\mathtt{TIME\_OF\_DAY}$, $\mathtt{DATE}$ and $\mathtt{DATE\_AND\_TIME}$) and 2) compound types ($\mathtt{ENUM}$ and $\mathtt{STRUCT}$). 
We implement these extended data types by the composition of built-in types and methods in $\mathbb{K}$ as follows.

\begin{figure}[!t]
\centering
\includegraphics[width=0.48\textwidth]{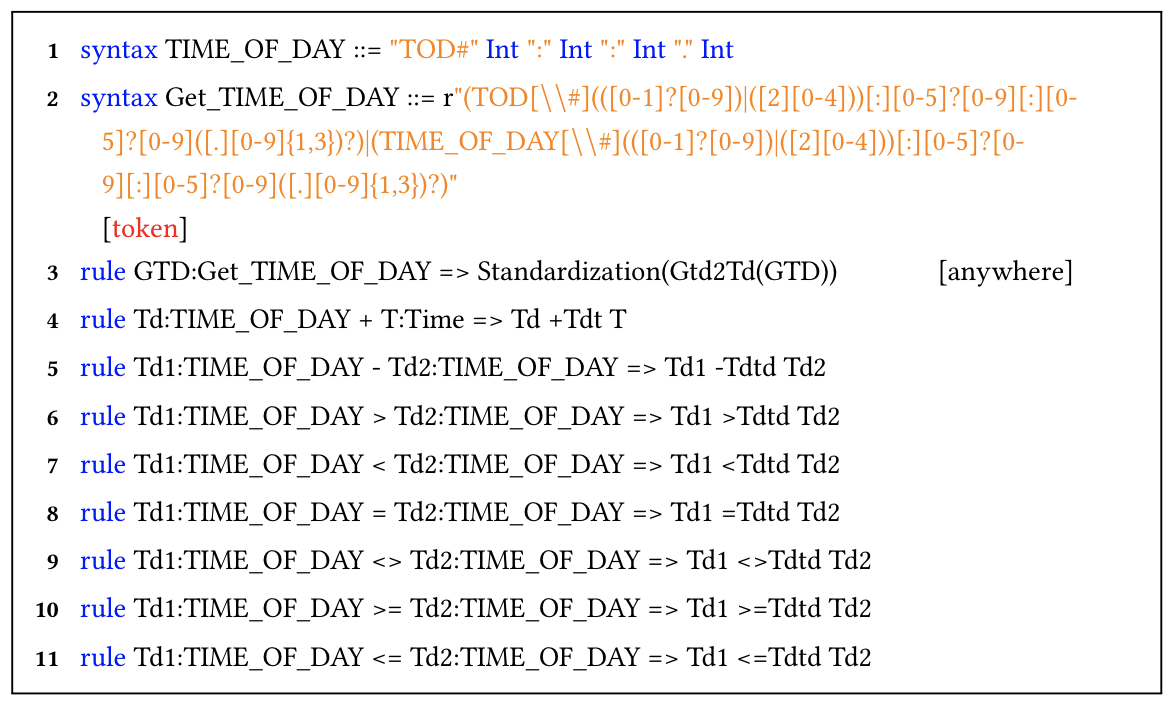}
\caption{Implementation of $\mathtt{TIME\_OF\_DAY}$ in $\mathbb{K}$} \label{fig5}
\end{figure}

We take $\mathtt{TIME\_OF\_DAY}$ as an example to introduce elementary types. 
There are two types of $\mathtt{TIME\_OF\_DAY}$ in ST, e.g., $\mathtt{TIME\_OF\_DAY\#23:45:56.30}$ and $\mathtt{TOD\#23:45:56.30}$. Fig.~\ref{fig5} shows our implementation of $\mathtt{TIME\_OF\_DAY}$ type together with its relevant operations.
Lines 1 and 2 respectively define the syntax of $\mathtt{TIME\_OF\_DAY}$ and how to parse it ($\mathtt{Get\_TIME\_OF\_DAY}$).
Line 3 is used to convert $\mathtt{Get\_TIME\_OF\_DAY}$ to $\mathtt{TIME\_OF\_DAY}$, which is achieved by two steps---$Gtd2Td$ and $Standardization$---where $Gtd2Td$ realizes the conversion of the format and $Standardization$ realizes content conversion, e.g., replacing $60$ minutes with $1$ hour. Lines 4--11 define some arithmetic and relational operations of $\mathtt{TIME\_OF\_DAY}$. 

\begin{figure}[!t]
\centering
\includegraphics[width=0.48\textwidth]{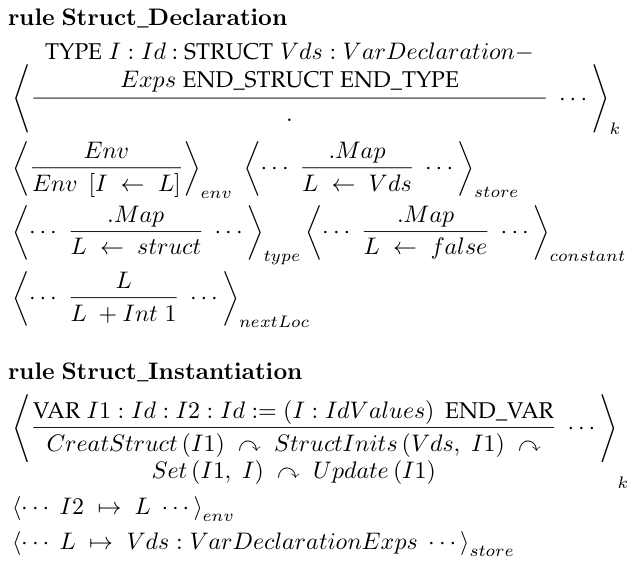}
\caption{The partial semantics of $\mathtt{STRUCT}$ in $\mathbb{K}$} \label{fig11}
\end{figure}

For compound types, we take $\mathtt{STRUCT}$ as an example and show its semantics in Fig.~\ref{fig11}, including both $\mathtt{STRUCT}$ declaration and instantiation.
Declarations are shown in $\mathbf{rule\; Struct\_Declaration}$, where we allocate memory for each defined data structure.
{The instantiation of $\mathtt{STRUCT}$ consists of four main steps in $\mathbf{rule\;Struct\_Instantiation}$}: 1)  $CreatStruct$ allocates memory for $I1$, 2) $StructInits$ generates each variable in turn according to $Vds$ in $\mathtt{STRUCT}$, 3) $Set$ assigns values to the corresponding variables according to $Idvs$, and finally, 4) $Update$ stores the mapping relationship of variables related to $I1$ into the memory of $I1$ to facilitate subsequent use.

\subsubsection{Main Control Statements} 
\begin{figure}[!t]
\centering
\includegraphics[width=0.48\textwidth]{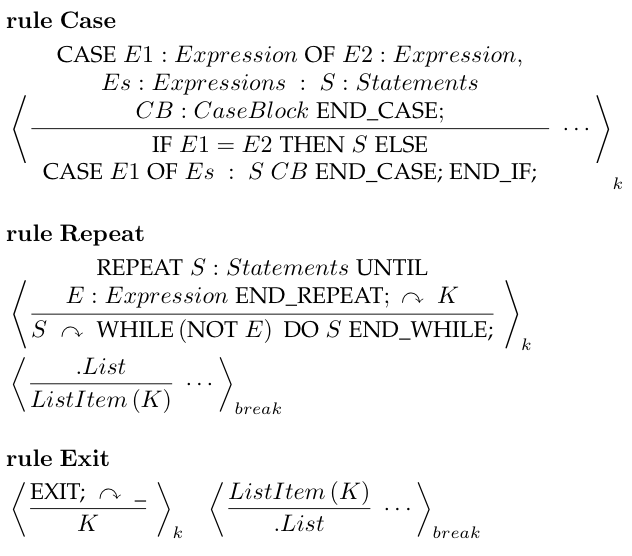}
\caption{The partial semantics of ST control statements} \label{fig12}
\end{figure}

Control statements are important in ST for achieving complex program logic (as in most other programming languages). We show the rules for $\mathtt{CASE}$, $\mathtt{REPEAT}$ and $\mathtt{EXIT}$ in Fig.~\ref{fig12} (as the semantics of $\mathtt{IF}$, $\mathtt{WHILE}$ and $\mathtt{FOR}$ are typical). A $\mathtt{CASE}$ statement can be rewritten as a combination of an $\mathtt{IF}$ and $\mathtt{CASE}$ through $\mathbf{rule\;Case}$.
The {$\mathbf{rule\;Repeat}$} is implemented as follows. We first store the subsequent statements outside the loop (recorded as $K$) in cell $break$ to deal with the $\mathtt{EXIT}$ statement that may appear, and then rewrite it into the form of $\mathtt{WHILE}$ for further execution. 
During the execution of the loop body, once $\mathtt{EXIT}$ is executed, all the statements in the current cell $k$ are discarded and rewritten to $K$ (storing the subsequent statements), as shown in $\mathbf{rule\;Exit}$.

\subsubsection{The Declaration and Call of POUs}
\begin{figure}[!t]
\centering
\includegraphics[width=0.48\textwidth]{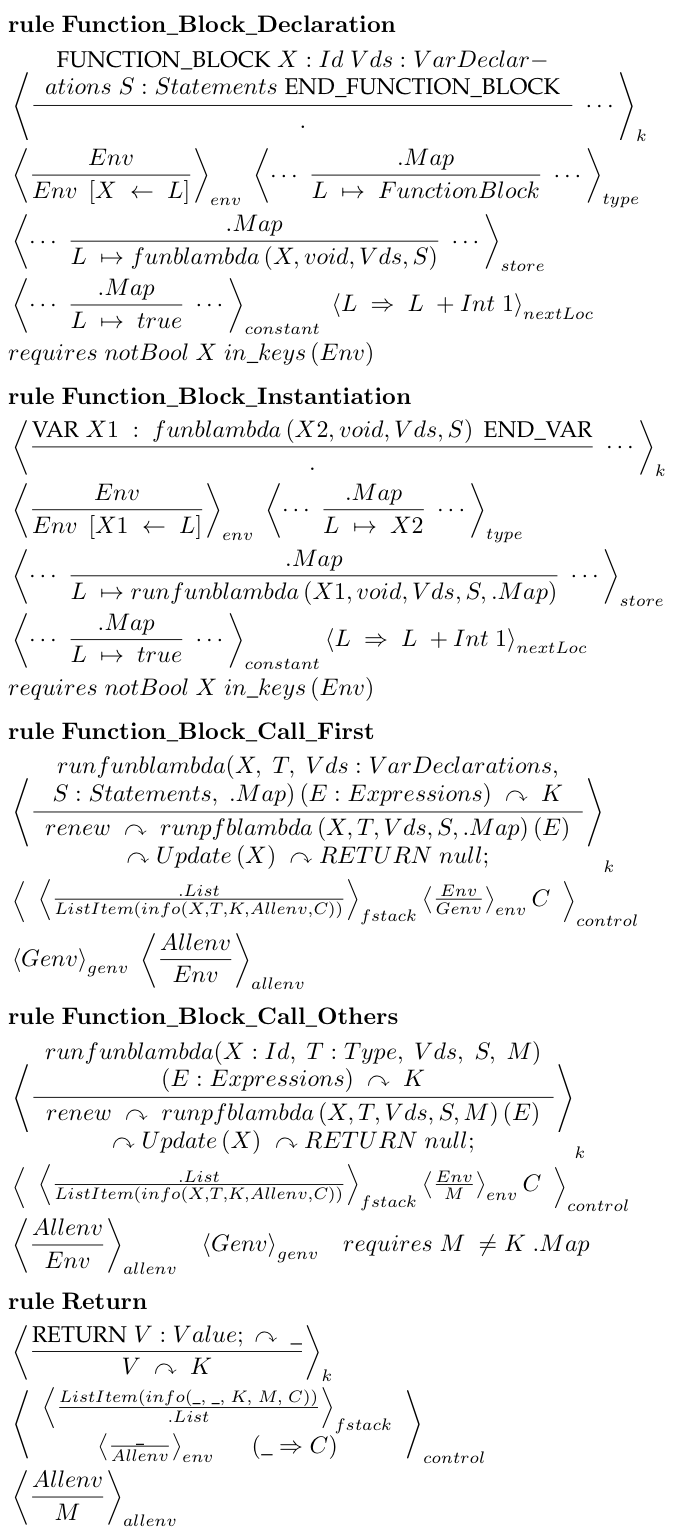}
\caption{The partial semantics of $\mathtt{FUNCTION\_BLOCK}$} \label{fig15}
\end{figure}

\begin{figure}[!t]
\centering
\includegraphics[width=0.48\textwidth]{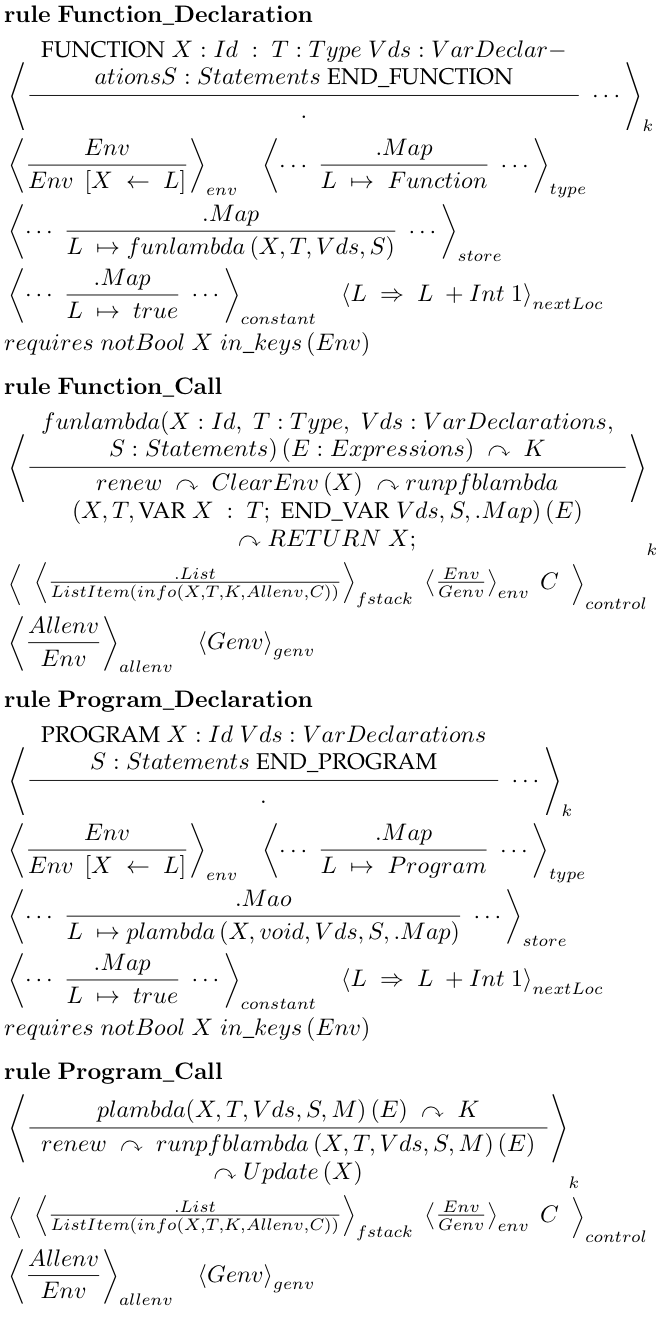}
\caption{The partial semantics of $\mathtt{FUNCTION}$ and $\mathtt{PROGRAM}$} \label{fig16}
\end{figure}

In ST programs, statements are inside Program Organization Units (POUs), i.e., $\mathtt{FUNCTION}$, $\mathtt{FUNCTION\_BLOCK}$ or $\mathtt{PROGRAM}$. A $\mathtt{FUNCTION}$ is a stateless POU type, comparing to a $\mathtt{FUNCTION\_BLOCK}$ which stores its own state after execution.  
The design of the $\mathtt{FUNCTION\_BLOCK}$ is similar to the concept of class-object manifestation in object-oriented programming (OOP), which aims to achieve better modularization.
$\mathtt{FUNCTION\_BLOCK}$s exist in two forms: as a type or as an instance, and only the instance can be called. For a $\mathtt{FUNCTION\_BLOCK}$ instance, the local variables retain their values between each `call'. $\mathtt{PROGRAM}$s are defined by the IEC 61131-3 standard as a ``logical assembly of all the programming language elements and constructs necessary for the intended signal processing required for the control of a machine or process by a PLC-system'' \cite{tiegelkamp2010iec}. Due to space constraints, we show the declaration, call and return operation of $\mathtt{FUNCTION\_BLOCK}$s in Fig.~\ref{fig15} as an example for illustration ($\mathtt{FUNCTION}$ and $\mathtt{PROGRAM}$ are shown in Fig.~\ref{fig16} and explained only when necessary).

\textbf{\textit{Declaration.}} The declaration of $\mathtt{FUNCTION\_BLOCK}$ is similar to that of $\mathtt{STRUCT}$. As shown in $\mathbf{rule\;Function\_Block\_Declaration}$, we first assign an index in memory for $\mathtt{FUNCTION\_BLOCK}\; X$, set the $type$ to the built-in $FunctionBlock$, and convert the entire declaration statement to the built-in type {$funblambda(X, void, $ $Vds, S)$} for storage, where $void$ means no return value, $Vds$ and $S$ are variable declarations and operations in $X$ respectively. The purpose of setting $const$ to $true$ is to prevent it from being modified. Note that $\mathtt{FUNCTION}$ and $\mathtt{PROGRAM}$ set $type$ and $store$ to $Function$, {$funblambda(X, T, Vds, S)$} and $Program$, {$plambda(X, void, Vds, S, .Map)$}.

\textbf{\textit{Instantiation.}} The instantiation of $\mathtt{FUNCTION\_BLOCK}$ is achieved through variable declarations, as shown in $\mathbf{rule\;Function\_Block\_Instantiation}$. However, the value is set to {$runfunblambda(X, void, Vds, S, .Map)$} to distinguish it from {$funblambda$} and $.Map$ is designed to store the $\mathtt{FUNCTION\_BLOCK}$ environment for next call and external query. This is because a $\mathtt{FUNCTION\_BLOCK}$ can only be called after instantiation, i.e., $runfunblambda$ can be executed but $funblambda$ can not. Since $\mathtt{FUNCTION}$ and $\mathtt{PROGRAM}$ have no such restrictions, {$funlambda$} and $plambda$ can be directly called and executed.

\textbf{\textit{Call.}} There are two cases when a $\mathtt{FUNCTION\_BLOCK}$ is called. The first case is that the $\mathtt{FUNCTION\_BLOCK}$ is called for the first time, as shown in $\mathbf{rule\;Function\_Block\_Call\_First}$. Since there is no initial environment (the last value of $runfunblambda$ is $.Map$), we will first store the current execution environment $info$ in $fstack$, including subsequent statements $K$, the $Allenv$ of the current environment, and the parameters $C$ in cell $control$. Then, we reset parameters $C$ through $renew$. After that, $\mathbb{K}$ executes the variable declaration $Vds$ (including index application, initialization and assignment) and statements $S$ in the function block. In addition, $Update$ is used to update the $.Map$ in {$runfunblambda$} to record the current environment. Finally, $\mathtt{RETURN}$ can return to the calling program and configure the corresponding environment. In other cases (not called for the first time), as shown in $\mathbf{rule\;Function\_Block\_Call\_Others}$, there is already a mapping relationship between related variables and values in cell $store$, and the mapping relationship between identifiers and indexes is also stored in the {$runfunblambda$}. Therefore, no new memory allocation will be made during the execution process and the existing environment will be used. Note that the value of the variable in the $\mathtt{FUNCTION\_BLOCK}$ will not be initialized, which means that the execution result for the same input may be different.

Regardless of whether $\mathtt{RETURN}$ appears in the $\mathtt{FUNCTION\_BLOCK}$, we add a $\mathtt{RETURN}$ by default for each $\mathtt{FUNCTION\_BLOCK}$ as a sign that the $\mathtt{FUNCTION\_BLOCK}$ has finished running and returned to the calling POUs. Since $\mathtt{FUNCTION\_BLOCK}$s and $\mathtt{PROGRAM}$s do not have a return value, we set $null$ as the return value. Note that a $\mathtt{FUNCTION}$ has a return value, and the returned value is the value corresponding to the function identifier, so we need to use $Clearenv$ to clean up the memory environment corresponding to the function identifier after calling procedure $renew$ and add the declaration of the function identifier variable in $Vds$. 

\subsubsection{Memory Operations}
Here, we present the rules for memory operations on elementary types in ST, such as built-in types and extended elementary types. What elementary types have in common is that they take only one memory slot. For complex types, such as enums, structs, arrays, etc, which are compositions of elementary types, the memory operation can be regarded as a set of memory operations on elementary types. For instance, the assignment to struct can be equivalent to assign value for each variable of this struct.

\begin{figure}[!t]
\centering
\includegraphics[width=0.48\textwidth]{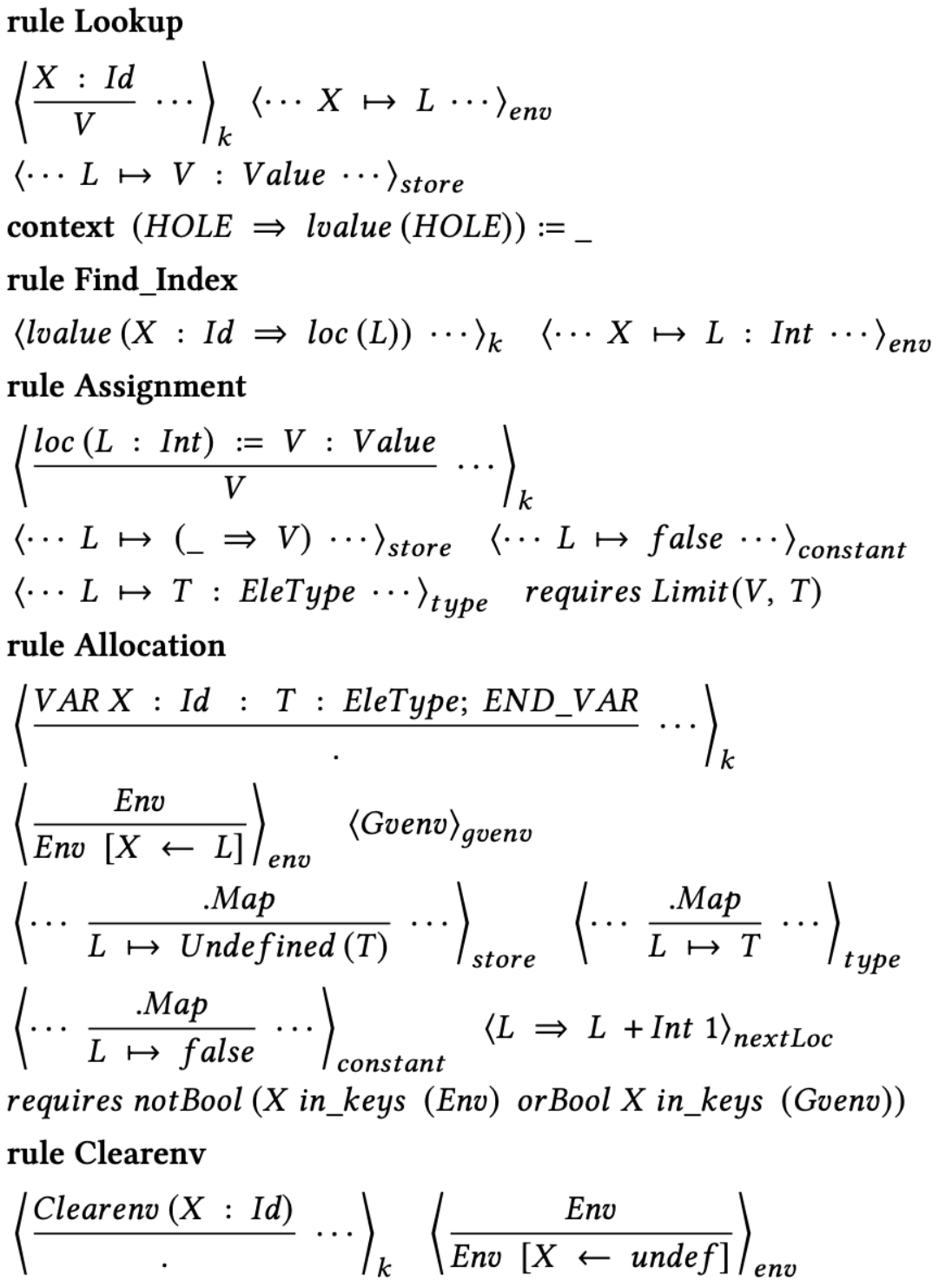}
\caption{The partial semantics of memory operations} \label{fig13}
\end{figure}

Similar to ST$_{demo}$, main memory operations in ST are still composed of $Allocation$, $Lookup$, $Assignment$ and additional $Clearenv$. Where $Allocation$ implements the allocation of memory for variables in the $store$, $Lookup$ is used to find variable values in $store$ cell, $Assignment$ implements the assignment of variables, and $Clearenv$ implements the recovery of memory in the $store$. However, because the complete ST semantics has a more complex type design, they will involve more cells in configurations, and are more complicated, as shown in Fig.~\ref{fig13}. 

Note that $HOLE$ is just a variable, but it has special meaning in the context of sentences with the `heat' or `cool' attribute. In short, `heat' is to lookup the corresponding content of the $HOLE$ in the formula, and `cool' is to put the recheck results back into the formula. For example, in expression $a + b$ where $a$ is represented as $HOLE$, `heat' is to take $a$ out of the formula and find its corresponding value. If it is 3, and `cool' puts 3 back into the original formula, then the formula becomes $3 + b$.

Let us start with the $Assignment$ operation (we omit $Lookup$ as it is straightforward). The $Assignment$ of ST divides the $Assignment$ of $\text{ST}_{demo}$ into two steps, where $\mathbf{context}$ and $\mathbf{rule \; Find\_Index}$ are used to determine the index $L$ of the assigned variable $X$ in $store$, and $\mathbf{rule \; Assignment}$ implements the update of the $store$ at index $L$. The purpose of this division is to make the $Assignment$ operation better applicable to complex types, because in some cases the index of the assigned variable can not be directly obtained and multiple queries are required. For instance, when assigning a value to $A \left[3,5,7\right]$, where $A$ is a multi-dimensional array, we need to look up each dimension one by one to finally determine the index. In addition, we refer to the state of $X$ in $type$ and $constant$ during the assignment process. On the one hand, we use $Limit$ to ensure that the assigned value meets the type requirements, and on the other hand, {we use $constant$ to ensure that the constant cannot be modified.}
Although the memory cleaning operation is not necessary for ST in $\mathbb{K}$, a simple $Clearenv$ operation can effectively reduce repetitive code and improve code readability. For $\mathbf{rule\;Clearenv}$, what needs attention is the operation on cell $env$: it replaces the index $L$ of variable $X$ with $undef$ which means null in the map supported by $\mathbb{K}$.

\begin{figure}[!t]
\centering
\includegraphics[width=0.48\textwidth]{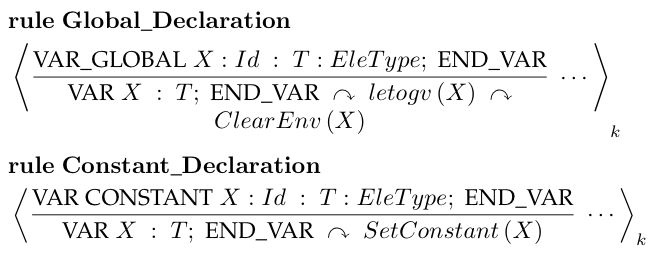}
\caption{The partial semantics of variable declarations} \label{fig14}
\end{figure}

ST has relatively complex and strict type definitions, therefore the $\mathbf{rule\;Allocation}$ of ST involves more cells and operations, such as $type$ and $constant$ for storing variable types and whether they are constants, where $Undefined$ is used to generate the default of the specified type. In addition, according to the content in TABLE~\ref{tab:element}, not only $\mathtt{VAR}$ will be used in the variable declaration process, but also other keywords, such as $\mathtt{VAR\_INPUT}$, $\mathtt{VAR\_IN\_OUT}$, etc. In order to reduce the complexity of the code, we also implement these declarations through $\mathtt{VAR}$ declarations. For instance, Fig.~\ref{fig14} shows the implementation of $\mathtt{VAR\_GLOBAL}$ and $\mathtt{CONSTANT}$. We realize regional changes (from the $env$ cell to the $gvenv$ cell) through $letogv$, and $SetConstant$ realizes the modification of the value in the $const$ cell.

We remark that K-ST covers 259 core features with 876 rules in total, using 2315 lines of $\mathbb{K}$ code. The complete code can be accessed through \url{https://github.com/wkyml/K-ST}. It has also been included by the $\mathbb{K}$ team in their projects, which can be found at the following link: \url{https://github.com/runtimeverification/k/blob/master/web/pages/projects.md}.

\section{Testing and Analysing ST Compilers}
\label{sec:applications}

In addition to providing formal references for defined languages, our formal semantics also has several applications that use language-independent tools provided by $\mathbb{K}$, such as state space exploration, model checking, symbol execution and deductive program validation. We omit demonstration of these applications in this paper since they have been well-illustrated in related works~\cite{park2015kjs,wang2018krust}. In this work, we introduce the testing of ST implementations/compilers based on our executable semantics, K-ST.

As discussed earlier, because ST compilers are typically provided by vendors, the execution behavior of compilers may be different, and may even be inconsistent with respect to the high-level semantics \cite{schumi2021spectest}. One of the main applications of the proposed semantics is to define the `reference' execution behavior of ST, which can help programmers detect bugs in existing ST compilers.

\begin{figure*}[!t]
\centering
\includegraphics[width=0.9\textwidth]{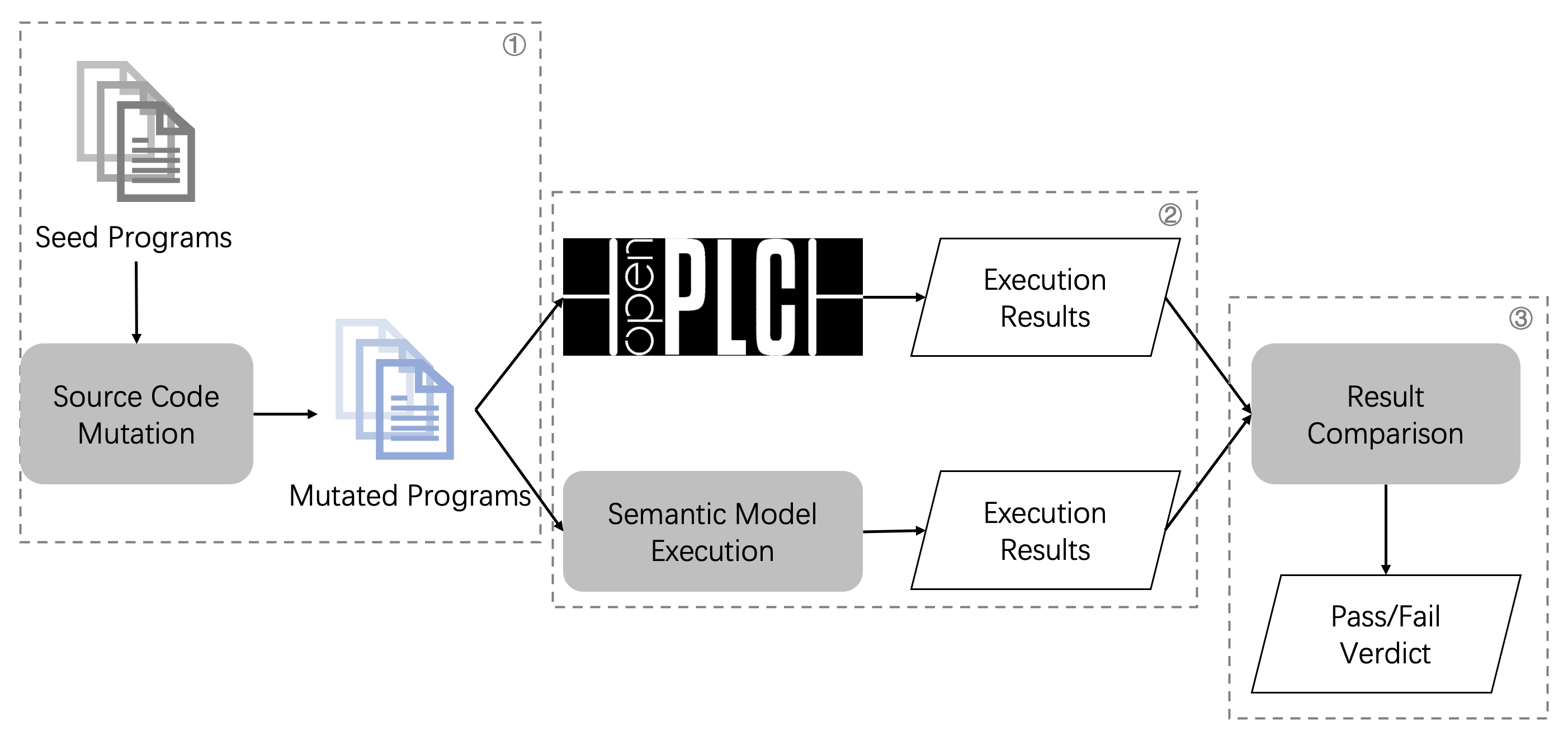}
\caption{Overview of the test process} \label{fig3}
\end{figure*}

To explore this application (and given the closed nature of commercial compilers), we choose OpenPLC\footnote{\url{https://www.openplcproject.com/}} as our test object, which is open source and supports ST programming. The overall workflow of our testing approach is depicted in Fig.~\ref{fig3}. It includes three parts: program variation, program execution and result comparison. First, seed programs are mutated to improve the diversity of test samples. Next, we use the mutated program as input to run OpenPLC and our executable semantics respectively. Finally, the result comparison part compares the consistency of the two execution results. It should be noted that we use a policy similar to \cite{le2014compiler}, that is, the program does not need input, and the category of result consistency comparison includes the values of all variables in the program. The comparison of results is performed to analyze potential inconsistencies between K-ST and OpenPLC.
By comparing the final execution state of the program with its variable state, we can identify potential inconsistencies. The execution state focuses on determining whether the program has completed its execution or terminates at the same statement. On the other hand, the variable state captures the values of all variables in the program, including input, output, and intermediate variables, after the program has finished running. TABLE 4 shows our measure of consistency, where $Q$ and $Q'$ represent the values of each variable after the program executes, $I$ and $I'$ represent the commands corresponding to the exception termination, and \CheckmarkBold and \XSolidBrush represent consistency and inconsistency respectively. As a result, unless K-ST and OpenPLC exhibit identical execution and memory states, their behavior will be deemed inconsistent.

\begin{table}[!t]
    \newcommand{\tabincell}[2]{\begin{tabular}{@{}#1@{}}#2\end{tabular}}
    \centering
    \begin{scriptsize}
    \caption{{Measure of K-ST/OpenPLC consistency}}
    \renewcommand\arraystretch{1.5}
    \begin{tabular}{cccc}
    \toprule
         \multicolumn{2}{c}{}  & \multicolumn{2}{c}{{The result of K-ST}} \\
         \cline{3-4}
         \multicolumn{2}{c}{} & \tabincell{c}{{Successful}\\{execution($Q$)}} & \tabincell{c}{{Unusual}\\{termination($I$)}} \\
         \hline
         & {Successful} & {$Q = Q'$ \CheckmarkBold} & \multirow{2}{*}{{\XSolidBrush}}\\
         {The result} & {execution($Q'$)} & {$Q \neq Q'$ \XSolidBrush} & \\
         \cline{2-4}
         {of OpenPLC} & {Unusual} & \multirow{2}{*}{{\XSolidBrush}} & {$I = I' \& Q = Q'$ \CheckmarkBold}\\
         & {termination($I'$)} & & {$others$ \XSolidBrush} \\
    \bottomrule
    \end{tabular}
    \label{tab:consistency}
    \end{scriptsize}
\end{table}

In order to better mutate seed programs to improve the diversity of test samples, we propose specific mutation operations in TABLE~\ref{tab:mutation} to generate mutated test samples. These mutation operations can enrich the test samples while minimizing program errors. Our method for generating mutant ST programs is shown in Algorithm~1. Given an ST program $S_i$, the algorithm makes a copy, randomly assigns initial values to all variables at the time of declaration, and applies some applicable mutation operators to randomly selected lines in the program. The test is done by comparing results of these samples in K-ST and OpenPLC. {It should be noted that correct and erroneous programs in the test sample are both meaningful for checking the consistency of execution behavior. This is because K-ST and OpenPLC report program errors at the same time, allowing us to verify a stronger notion of consistency.} In addition, considering the lag of OpenPLC updates, we also tested it on the latest Beremiz\footnote{\url{https://beremiz.org/}} which uses the same underlying implementation (MATIEC \footnote{\url{https://github.com/thiagoralves/OpenPLC\_Editor/tree/master/matiec}}) as OpenPLC.
The specific results of the test are shown in Section~\ref{section5}.

\begin{table}[!t]
    \centering
    \caption{Mutation operations}
    \renewcommand\arraystretch{1.5}
    \begin{tabular}{c|c}
    \toprule
      Mutation Operation &  Example\\
     \hline
      Variable Random Assignment  &  $a : INT; \rightsquigarrow a : INT := 3527;$\\
      Scalar Variable Replacement & $a := b; \rightsquigarrow a := c\ |\ 30;$\\
      Arithmetic Operator Replacement & $a + b \rightsquigarrow a - b$\\
      Arithmetic Operator Insertion & $a + b \rightsquigarrow a + b - c$\\
      Arithmetic Operator Deletion & $a + b - c \rightsquigarrow a + b$\\
      Relational Operator Replacement & $a > b \rightsquigarrow a <= b$\\
      Logical Connector Replacement & $a\ AND\ b \rightsquigarrow a\ OR\ b$\\
      Logical Connector Insertion & $a\ AND\ b \rightsquigarrow a\ AND\ b\ OR\ c$\\
      Logical Connector Deletion & $a\ AND\ b\ OR\ c \rightsquigarrow a\ AND\ b$\\
      ``NOT'' Mutation & $NOT\ a \rightsquigarrow a\ |\ a \rightsquigarrow NOT\ a$\\
      Statement Insertion & $\ \rightsquigarrow IF \cdots END\_IF;$\\
      Statement Deletion & $EXIT; \rightsquigarrow \ $\\
    \bottomrule
    \end{tabular}
    \label{tab:mutation}
\end{table}

\begin{figure}
	\centering
\includegraphics[width=\linewidth]{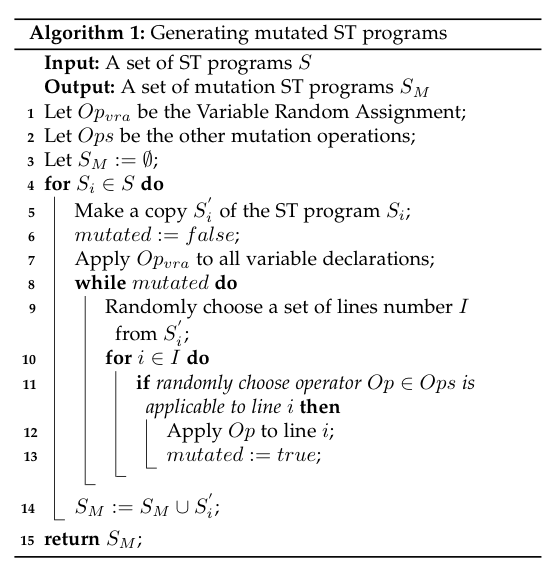}
\end{figure}

\section{Evaluation}
\label{section5}\label{sec:evaluation}
In order to evaluate the semantics of ST we defined in $\mathbb{K}$, we deployed K-ST on $\mathbb{K}$ version 5.1.11 (Intel(R) Core(TM) i7-9750H CPU @ 2.60GHz). In the following, we design multiple experiments to systematically answer the following research questions (\emph{RQs}).
\begin{itemize}
    \item \emph{RQ1: How much of the ST language is K-ST covering?} Completeness of the semantics is an important indicator to measure executable formal semantics. The lack of key semantics will seriously affect the usefulness of formal semantics.
    \item \emph{RQ2: Is K-ST correct?} Semantic correctness is the basis for ensuring the usability of executable formal semantics, so we need to analyze the correctness of formal semantics implemented.
    \item \emph{RQ3: Can K-ST be used to discover bugs in a compiler?} This is important since a key application of executable formal semantics is to identify compiler bugs.
\end{itemize}

\subsection{Test Sets}
For the purpose of evaluating the coverage and the correctness of K-ST, the test data set that we used comes from GitHub. We searched 4853 programs in GitHub through keywords in the ST language. Then, we automatically screened out samples containing other programming languages (2516) and XML forms (1542). After that, we manually splice the remaining programs and remove samples that lack the components required for operation (such as POUs). After screening, 567 complete programs written in pure ST formed our test set. In other words, these 567 samples contain all the components required for operation and do not use other languages, such as C and Python.

With the aim of comprehensively testing the correctness of the execution behavior of OpenPLC, we use two sample sets, including test samples collected from GitHub (GitHub set) and test samples obtained through mutation (Mutated set). The GitHub set is the sample set with 567 test samples mentioned before. The Mutated set is generated by Algorithm 1. We selected 30 high-quality samples from GitHub set as initial mutant seeds. These 30 samples contain all the key features of ST. Then, three rounds of iterative mutation are carried out through Algorithm 1. Each round of iteration produces 10 mutation samples per seed. Except for the initial seed used in the first round, the seeds of each round of mutation are the result of the previous round of mutation. We get a set containing 33,330 mutation samples.

\subsection{Experiment Results and Analyses}
\subsubsection{Semantic Completeness (RQ1)}
\begin{figure*}
\centering
\includegraphics[width=\textwidth]{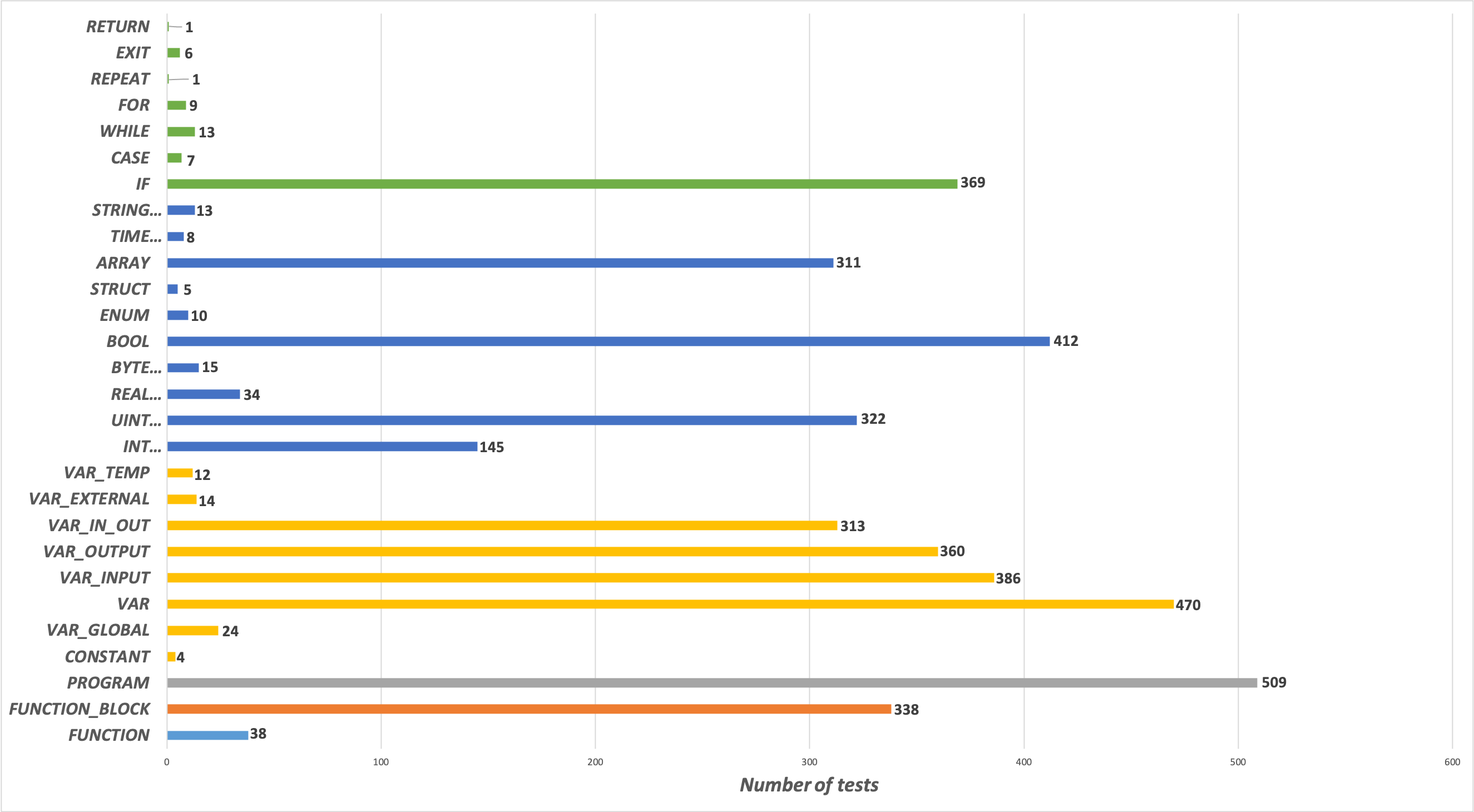}
\caption{Number of tests for each feature in ST} \label{fig6}
\end{figure*}

We executed K-ST on 567 test samples collected from GitHub. Among these 567 test samples, K-ST supports the execution of 509 of them. For these 509 tests which K-ST can support, Fig.~\ref{fig6} lists the number of tests for some important features (based on TABLE~\ref{tab:element}) used in the evaluation. Specifically, there are six kinds of features, namely $\mathtt{FUNCTION}$, $\mathtt{FUNCTION\_BLOCK}$, $\mathtt{PROGRAM}$, $Declaration \; types$, $Date \; types$ and $Statements$. For $Declaration$ $types$, we list the number of tests for $\mathtt{CONSTANT}$, $\mathtt{VAR\_GLOBAL}$, $\mathtt{VAR}$, $\mathtt{VAR\_INPUT}$, $\mathtt{VAR\_OUTPUT}$, $\mathtt{VAR\_IN\_OUT}$, $\mathtt{VAR\_TEMP}$ and $\mathtt{VAR\_EXTERNAL}$. For $Data \; types$, we list the number of tests for elementary types signed integer ($\mathtt{INT}$, $\mathtt{DINT}$, $\mathtt{SINT}$, $\mathtt{LINT}$), unsigned integer ($\mathtt{UINT}$, $\mathtt{UDINT}$, $\mathtt{USINT}$, $\mathtt{ULINT}$), float ($\mathtt{REAL}$, $\mathtt{LREAL}$), Boolean ($\mathtt{BOOL}$), byte ($\mathtt{BYTE}$, $\mathtt{WORD}$, $\mathtt{DWORD}$), string ($\mathtt{STRING}$, $\mathtt{WSTRING}$), and time ($\mathtt{TIME}$, $\mathtt{DATE}$, $\mathtt{TIME\_OF\_DAY}$, $\mathtt{DATE\_AND\_TIME}$); compound types enum ($\mathtt{ENUM}$) and struct ($\mathtt{STRUCT}$); and finally, the array type $\mathtt{ARRAY}$. For $Statements$, we list the number of tests for main control statements: $\mathtt{IF}$, $\mathtt{CASE}$, $\mathtt{FOR}$, $\mathtt{WHILE}$, $\mathtt{REPEAT}$, $\mathtt{EXIT}$ and $\mathtt{RETURN}$. 

As indicated in Fig.~\ref{fig6}, compared with $\mathtt{FUNCTION}$, the $\mathtt{FUNCTION\_BLOCK}$ is more favored by ST programmers ($\mathtt{PROGRAM}$ is necessary for ST program operation). For $Declaration \; types$, the most used is $\mathtt{VAR}$ (with a ratio of $470/509$), followed by $\mathtt{VAR\_INPUT}$ ($386/509$), $\mathtt{VAR\_OUTPUT}$ ($360/509$) and $\mathtt{VAR\_IN\_OUT}$ ($313/509$). Among all the $Data \; types$, $\mathtt{BOOL}$ is the most used, followed by unsigned integer and $\mathtt{ARRAY}$, constituting $322/509$ and $311/509$ respectively. For the $Data \; types$, $\mathtt{BOOL}$ is the most common type. In addition, we must remark that we do not count the type of array members. Finally, $\mathtt{IF}$ is the most common statement in all the tests considered. This is also in line with the main working scenarios of PLCs. 

We remark that we do not consider the vendor-based functions in this experiment as these functions vary not only from vendor to vendor, but even from product to product. In particular, Mitsubishi PLCs provide completely different data types, including Bit, Word[Signed/Unsigned], Double Word[Signed/Unsigned], Bit STRING[16-bit/32-bit], FLOAT, STRING[32] and Time. Siemens PLCs support keyword $\mathtt{BEGIN}$ to represent the end of variable declaration and the beginning of operation instructions. In addition, there are also obvious differences between different products of the same vendor. For example, the S7-1500 and the S7-1200 from Siemens support different type conversion methods\footnote{\url{https://support.industry.siemens.com/dl/dl-media/272/109742272/att\_918238/v6/93516999691/zh-CHS/index.html\#ae443583b99950f7cca0d7237fe81ad4}}, where the former only provides explicit conversions of types, and the latter provides both explicit and implicit conversions.

\subsubsection{Semantics Correctness (RQ2)}
On the other hand, in order to evaluate the correctness of K-ST, we compared the execution results of K-ST against those of vendor compilers CODESYS, CX-Programmer and GX Works2. We consider the proposed semantics correct if the execution behaviors of K-ST are consistent with the ones of the CODESYS, CX-Programmer and GX Works2 compilers. 
The consistency criteria described in Section~\ref{sec:applications} are utilized to evaluate the consistency of behavior between K-ST and the compilers provided by vendors. Specifically, if K-ST and these compilers demonstrate identical execution and variable states for the same program, their behavior is deemed consistent.
We list the coverage of the K-ST semantics in TABLE~\ref{tab:cover} from the perspective of each feature specified by the official ST documentation, where FC, C and N mean ``Fully Covered and Consistent with Compilers'', ``Covered and Consistent with Compilers'' and ``Not Covered'', respectively.
\begin{table*}
  \caption{Coverage of the proposed ST semantics}
  \label{tab:cover}
  \renewcommand\arraystretch{1.5}
  \begin{tabular}{lclclc}
    \toprule
    Feature & Coverage & Feature & Coverage & Feature & Coverage\\
    \midrule
	$\mathbf{POUs(core)}$ & & $\mathbf{Data \; types(core)}$ & & $\;\;\;\;\; Enum \; instantiation$ & FC\\
	$POUs \; declaration$ & & $\;\;\;\;\;SINT$ & FC & $Struct$ & \\
	$\;\;\;\;\;FUNCTION$ & FC & $\;\;\;\;\;INT$ & FC & $\;\;\;\;\;Struct \; declaration$ & FC\\
	$\;\;\;\;\;FUNCTION\_BLOCK$ & FC & $\;\;\;\;\;DINT$ & FC & $\;\;\;\;\;Struct \; instantiation$ & FC \\
	$\;\;\;\;\;PROGRAM$ & FC & $\;\;\;\;\;LINT$ & FC & $Function \; block$ &\\
	$POUs \; calls$ &  & $\;\;\;\;\;USINT$ & FC & $\;\;\;\;\;Function \; block \; instantiation$ & FC\\
	$\;\;\;\;\;FUNCTION$ & FC & $\;\;\;\;\;UINT$ & FC & $Array$ & \\
	$\;\;\;\;\;FUNCTION\_BLOCK$ & FC & $\;\;\;\;\;UDINT$ & FC & $\;\;\;\;\;One-dimensional \; array$ & C \\
	$\;\;\;\;\;PROGRAM$ & FC & $\;\;\;\;\;ULINT$ & FC & $\;\;\;\;\;Multi-dimensional \; array$ & C \\
	$\mathbf{Variable \; Declaration(core)}$ & & $\;\;\;\;\;REAL$ & FC & $\mathbf{Statements(core)}$ &  \\
	$\;\;\;\;\;CONSTANT$ & FC & $\;\;\;\;\;LREAL$ & FC & $Assignment \; statement$ & \\
	$\;\;\;\;\;VAR\_GLOBAL$ & FC & $\;\;\;\;\;BOOL$ & FC & $\;\;\;\;\;:=$ & FC\\
	$\;\;\;\;\;VAR$ & FC & $\;\;\;\;\;BYTE$ & FC & $\;\;\;\;\;\Rightarrow$ & N\\
	$\;\;\;\;\;VAR\_INPUT$ & FC & $\;\;\;\;\;WORD$ & FC & $Branch \; statement$ &\\
	$\;\;\;\;\;VAR\_OUTPUT$ & FC & $\;\;\;\;\;DWORD$ & FC & $\;\;\;\;\;IF$ &FC \\
	$\;\;\;\;\;VAR\_IN\_OUT$ & FC & $\;\;\;\;\;STRING$ & FC & $\;\;\;\;\;CASE$ &FC\\
	$\;\;\;\;\;VAR\_EXTERNAL$ & FC & $\;\;\;\;\;WSTRING$ & FC & $Loop \; statement$ & \\
	$\;\;\;\;\;VAR\_TEMP$ & FC & $\;\;\;\;\;TIME$ & FC & $\;\;\;\;\;WHILE$ &FC\\
	$\;\;\;\;\;AT$ & C & $\;\;\;\;\;DATE$ & FC & $\;\;\;\;\;FOR$ & FC\\
	$\;\;\;\;\;RETAIN$ & N & $\;\;\;\;\;TIME\_OF\_DAY$ & FC & $\;\;\;\;\;REPEAT$ & FC\\
	$\;\;\;\;\;PERSISTENT$ & N & $\;\;\;\;\;DATE\_AND\_TIME$ & FC & $Break \; statement$ &\\
	$\mathbf{Typed \; constant}$& & $Enum$ & & $\;\;\;\;\;RETURN$ &FC\\
	$\;\;\;\;\;\mathbf{Type} \; \# \; \mathbf{Data}$ & FC & $\;\;\;\;\; Enum \; declaration$ & FC & $\;\;\;\;\;EXIT$ &FC\\
	\hline
	$\mathbf{Built-in \; function}$& & & & &\\
	$Numerical \; function \; \left( 30 \right)$& & & & &\\
	\multicolumn{6}{c}{$ADD$, $SUB$, $MUL$, $SQR$, $INC$, $DEC$, $MAX$, $MIN$, $MUX$, $ABS$, $SQRT$, $TRUNC$, $FRAC$, $FLOOR$, $LN$, $LOG$, $EXP$, $SIN$}\\
	\multicolumn{6}{c}{$COS$, $TAN$, $COS$, $TAN$ $ASIN$, $ACOS$, $ATAN$, $NEG$, $EXPT$, $DIV$, $MOD$, $LIMIT$}\\
	$Logical \; function \; \left( 9 \right)$& & & & &\\
	\multicolumn{6}{c}{$GT$, $LT$, $GE$, $LE$, $EQ$, $NE$, $AND$, $OR$, $SEL$}\\
	$String \; function \; \left( 9 \right)$& & & & &\\
	\multicolumn{6}{c}{$CONCAT$, $INSERT$, $DELETE$, $REPLACE$, $FIND$, $LEN$, $LEFT$, $RIGHT$, $MID$}\\
	$Translate \; function \; \left( 160 \right)$& & & & &\\
  \bottomrule
\end{tabular}\\
\begin{scriptsize}
\centering
$\square$\ FC: Fully Covered and Consistent with Compilers (256/262) \qquad $\square$\ C: Covered and Consistent with Compilers (3/262) \qquad $\square$\ N: Not Covered (3/262)
\end{scriptsize}
\end{table*}

From TABLE~\ref{tab:cover}, we can see clearly that for POUs, we fully cover the declaration and call. In variable declarations, $\mathtt{AT}$ is related to input and output. We remark, however, that the storage mode of variables in $\mathbb{K}$ is very different from that in real PLCs, so we just support simple computer-side input and output. In addition, $\mathtt{RETAIN}$ and $\mathtt{PERSISTENT}$ are related to the actual situation in the PLC, so they are not implemented. For instance, $\mathtt{AT}$ is used to bind the actual point of the PLC; $\mathtt{RETAIN}$ and $\mathtt{PERSISTENT}$ support the preservation of variable values after a power failure or power loss. $Array$ is the only one which is covered but not fully covered in all data types. Limited by the realization of arrays, it is temporarily impossible to achieve the array for enum and struct, and to assign values to multi-dimensional arrays as a whole. In statements, $\Rightarrow$ has been used in $\mathbb{K}$ and can be replaced by $:=$. For built-in functions, we show a list which we supported, including 30 numerical functions, 9 logical functions, 9 string functions and 160 translate functions.

In the process of comparing with CODESYS, CX-Programmer and GX Works2, the following points need to be explained. Firstly, due to the closed nature of these compilers, they cannot be simply called, so we have to manually fill the code in the specified way into the compiler to compile and run, and compare the results, which is laborious and tedious work. This also hinders us from testing these commercial compilers in an extensively large scale. After that, different vendors have obvious differences in the implementation of compilers, so the source code needs to be adapted to a certain extent. For example, only 10 basic data types---Bit, Word[Signed/Unsigned], Double Word[Signed/Unsigned], Bit STRING[16-bit/32-bit], FLOAT, STRING[32] and Time---are provided in the GX Works2 compiler, so we need to adapt the variable types of the source program.

\subsubsection{Finding Bugs in OpenPLC (RQ3)}
We execute OpenPLC and K-ST with the GitHub set and Mutated set as input. The execution results of the two data sets are shown in TABLE~\ref{tab:result}. Here, $K_pO_f$ is the number of programs that K-ST can execute normally but OpenPLC cannot compile and run; $K_fO_p$ is the number of programs that K-ST cannot run normally but OpenPLC can.
\begin{table}[!t]
    \centering
    \caption{The results of K-ST and OpenPLC}
    \renewcommand\arraystretch{1.5}
    \begin{tabular}{c|c|c|c}
    \toprule
         \multicolumn{2}{c|}{Data Set} & GitHub Set & Mutated Set  \\
         \hline
         \multicolumn{2}{c|}{Number of samples} & 567 & 31059 (2271) \\
         \hline
         Number of program & K-ST & 509 & 15850 \\
         \cline{2-4}
         run completely & OpenPLC & 490 & 11581 \\
         \hline
         \multirow{3}{*}{Inconsistent} & $K_pO_f$ & 30 & 5664 \\
         \cline{2-4}
         & $K_fO_p$ & 11 & 1395 \\
         \cline{2-4}
         & Diff.~Result & 0 & 735 \\
         \bottomrule
    \end{tabular}
    \label{tab:result}
\end{table}

For the GitHub set, K-ST supports 509 of them, and OpenPLC supports 490. Through analysis, we found that the reason for this phenomenon is that OpenPLC has some functional deficiencies. For example, OpenPLC does not support the initialization of variables using formulas at the time of declaration; numerical calculations of $\mathtt{BYTE, WORD, DWORD}$ types are not supported, etc. 

For the Mutated set, there is a big difference between the execution results of K-ST and OpenPLC. First of all, we filter 2,271 timeout programs that timed out both in OpenPLC and K-ST with 10 seconds as the time limit. 
After that, we manually analyzed these samples with inconsistent results to determine the causes. For the large $K_pO_f$ value, functional deficiencies remain the main reason.

We found an interesting bug in OpenPLC. The bug is a ``VAR'' parsing exception in OpenPLC. If the first operation instruction starts with ``VAR'', such as ``$\text{VAR}0 := 1;$'', OpenPLC terminates abnormally. The interesting phenomenon is when an error statement appears in an unexecuted part of the program, such as after the "$\mathtt{RETURN;}$": K-ST can execute such a program, but OpenPLC cannot. The main reason for this phenomenon is that $\mathbb{K}$ adopts an operation-based detection mechanism. Because the error code will not be executed, it will not lead to the termination of our executable semantics. The case study is shown in APPENDIX~A.

After that, by analyzing those programs that have different results on K-ST and OpenPLC, we find that the reasons for the different results are mainly due to the differences in underlying implementations between $\mathbb{K}$ and OpenPLC. For example, for integer mode operation $-7\ \text{MOD}\ 3$, the execution result of K-ST is $-1$, whereas the result for OpenPLC is $2$. From a mathematical point of view, both results are correct, but they will have a completely different impact on any following operations. When we run the program again in CODESYS, the results of CODESYS were the same as K-ST. 

For those samples that K-ST cannot run normally but OpenPLC can execute normally, our analysis found some bugs in OpenPLC. For example, while OpenPLC can check explicit divide-by-zero operations, it allows the execution of implicit divide-by-zero operations. TABLE~\ref{tab:OpenPLC} details all functional deficiencies and bugs we found in OpenPLC. We show some relevant case studies in APPENDIX~B. Considering that Beremiz can be regarded as an updated version of OpenPLC, we have retested the inconsistencies we found in Beremiz. We found that in the latest Beremiz, it fixes some problems, including negative MOD operation results and ``VAR'' parsing exceptions. But other bugs and shortcomings still exist. In response to these problems in OpenPLC, we have submitted them to the OpenPLC and Beremiz developers and are waiting for their confirmation\footnote{\url{https://bitbucket.org/automforge/matiec\_git/issues?status=new \&status=open}}.
\newcommand{\tabincell}[2]{\begin{tabular}{@{}#1@{}}#2\end{tabular}}
\begin{table*}[!t]
    \centering
    \caption{The bugs and functional deficiencies of OpenPLC}
    \renewcommand\arraystretch{1.5}
    \begin{tabular}{c|c|c}
    \toprule
        Type & Problem & Description  \\
        \hline
        \multirow{5}{*}{Bug} & ``VAR'' parsing exception & The first operation instruction starts with ``VAR'', and OpenPLC terminates abnormally.\\
        \cline{2-3}
        & Division by zero & OpenPLC can check explicit division 0 but allow the execution of implicit division 0.\\
        \cline{2-3}
        & Overflow access & OpenPLC can check explicit overflow access but allow the execution of implicit overflow access.\\
        \cline{2-3}
        & MOD by zero & OpenPLC provides MOD 0 operation, and the result is 0.\\
        \cline{2-3}
        & MOD Exception & The divisor of MOD operation can be empty.\\
        \hline
        \multirow{9}{*}{\begin{tabular}[c]{@{}c@{}}Functional\\deficiencies\end{tabular}} & \multirow{2}{*}{\begin{tabular}[c]{@{}c@{}}Numerical\\calculation defects\end{tabular}} &  OpenPLC does not support normal numerical calculation $**$.\\
        \cline{3-3}
        & & Numerical calculations of $\mathtt{BYTE, WORD, DWORD}$ types are not supported.\\
        \cline{2-3}
        & Array functions defects & Parentheses are not allowed in array assignments.\\
        \cline{2-3}
        & \tabincell{c}{$\mathtt{FUNCTION\_BLOCK}$\\instantiation defects} & Multiple instantiation of function blocks in one statement is not allowed.\\
        \cline{2-3}
        & $\mathtt{ENUM}$ defects & OpenPLC does not support normal assignment of $\mathtt{ENUM}$ type.\\
        \cline{2-3}
        & \multirow{2}{*}{\begin{tabular}[c]{@{}c@{}}Variable\\declaration defects\end{tabular}} & Some non-keyword strings cannot be used as variable names, such as ``ramp'', ``LocalVar0\_'', etc.\\
        \cline{3-3}
        & & OpenPLC can not support formula and other variables previously declared as initial value.\\
        \cline{2-3}
        & \multirow{2}{*}{Structural defects} & Without operation or variable declarations, OpenPLC cannot compile ST program.\\
        \cline{3-3}
        & & Without statements in $\mathtt{FOR, WHILE, IF, CASE, REPEAT}$, OpenPLC cannot compile ST.\\
    \bottomrule
    \end{tabular}
    \label{tab:OpenPLC}
\end{table*}

\section{Related Work}
\label{sec:related_work}
In this section, we discuss some other PLC program analysis techniques, summarize their characteristics, and distinguish them from our work.

Keliris et al. \cite{keliris2018icsref} propose a framework (ICSREF) which can automate the reverse engineering process for PLC binaries. They instantiate ICSREF modules for reversing binaries compiled with CODESYS and getting the complete Control Flow Graph (CFG), and they provide an end-to-end case study of dynamic payload generation and attack deployment. Tychalas et al. \cite{tychalas2021icsfuzz} analyze the binary files generated by all control system programming languages in CODESYS to understand the differences and even the vulnerabilities introduced during the program compilation process. Based on this analysis, they provide a fuzzing framework (ICSFuzz) to perform security evaluation of the PLC binaries. Our work differs from them because we focus on the source code and do not rely on any specific compilation environment. 

Kuzmin et al. \cite{kuzmin2013construction} propose to use linear-time temporal logic (LTL) to guide program behavior and check whether ST programs satisfy the corresponding temporal logic through Cadence SMV. Darvas et al. \cite{darvas2014formal} propose rule-based reductions and a Cone of Influence (COI) reduction variant for state explosion problems that may be encountered in the formal analysis of ST code, and use the NuSMV model checker to verify temporal logic. After that, they \cite{darvas2015formal} provide a state machine and data-flow-based formal specification method for PLC modules. In addition, they \cite{darvas2016generic} analyze the feasibility of converting between the 5 PLC programming languages provided by Siemens, and point out that the extended SCL (a vendor-defined ST) can be used as the target language for conversion. Adiego et al. \cite{adiego2015applying} propose an intermediate model-based method which can transform PLC programs written in different modeling languages of verification tools to facilitate checking temporal logic. Hailesellasie et al. \cite{hailesellasie2018intrusion} propose UBIS, which converts ST programs with potential intrusions as well as trusted versions of programs into attributed graphs through UPPAAL, and compares their nodes and edges to detect stealthy code injections. Bohlender et al. \cite{bohlender2018compositional} apply formal verification and falsification of temporal logic specifications to analyze chemical plant automation systems. Rawlings et al. \cite{rawlings2018application} use symbolic model checking tools st2smv and SynthSMV to verify and falsify a ST program controlling batch reactor systems. Xiong et al. \cite{xiong2020safety} use the behavior model (BM) to specify the behavior of ST programs, and provide an method based on automatic theoretical to verify LTL attributes on BM. Our work differs from the aforementioned works because they attempt to transform PLC programs into intermediate languages or other programming languages which are suitable for verifying or detecting potential issues, and lack analysis in the conversion process. In addition, these methods do not offer feedback at the level of source code. 

Huang et al. \cite{huang2019kst} is the closest work to ours. They first defined the executable semantics of the ST language in $\mathbb{K}$ and use it to check some security properties. Our work differs because we cover a more complete ST language, and we can use it to discover errors in ST compilers.

\section{Conclusion}
\label{sec:conclusion}
In this paper, we introduced an executable operational semantics of ST formalized in the $\mathbb{K}$ framework. We presented the semantics of the core features of ST, namely data types, memory operations, its main control statements, and function calls. Our experimental results show that the proposed ST semantics has already covered the main core language features and correctly implements 26,137 lines of public ST code on GitHub. Furthermore, the application of the proposed semantics in testing and analyzing PLC compilers is discussed. By comparing and analyzing the execution results of OpenPLC and K-ST, we found five bugs and some functional deficiencies in OpenPLC.
In the future, we hope to further extend K-ST to support the programming environments provided by different vendors. For example, vendors may customize keywords ($\mathtt{Bit\;STRING}$ of GX Works2), add additional structures ($\mathtt{LABEL}$ of Siemens), or even widely extend ST (ExST of CODESYS).

\section*{Acknowledgments}

We thank the reviewers for their constructive feedback. This research is supported by National Key R\&D Program of China under grant 2020YFB2010900, NSFC under grants 61833015 and 62293511, Provincial Key R\&D Program of Zhejiang under grants 2020C01038 and 2021C01032, and the Starry Night Science Fund of Zhejiang University Shanghai Institute for Advanced Study, Grant No. SN-ZJU-SIAS-001.

\ifCLASSOPTIONcaptionsoff
  \newpage
\fi



\bibliographystyle{IEEEtran}
\bibliography{IEEEabrv}
%



%
\begin{IEEEbiography}[{\includegraphics[width=1in,height=1.25in,clip,keepaspectratio]{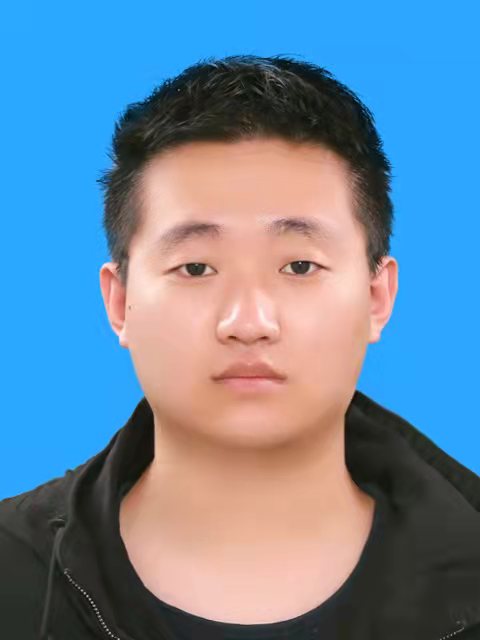}}]{Kun Wang} received the B.S. degree in information and computing sciences from Chongqing University of Posts and Telecommunications of China, in 2017. He received the M.Eng. degree in Cyberspace Security from Xidian University of China, in 2020. He is currently pursuing his Ph.D degree with State Key Laboratory of Industrial Control Technology, Group of Networked Sensing and Control, Zhejiang University. His research interests include control system security and formal methods.
 \end{IEEEbiography}

 \begin{IEEEbiography}[{\includegraphics[width=1in,height=1.25in,clip,keepaspectratio]{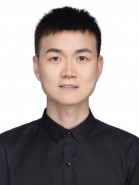}}]{Jingyi Wang} is currently a tenure-track assistant professor at the College of Control Science and Engineering, Zhejiang University, China. He received his Ph.D. from Singapore University of Technology and Design in 2018, and his bachelor’s degree in Information Engineering from Xi’an Jiaotong University in 2013. He was a research fellow at the School of Computing, National University of Singapore during 2019-2020 and at Information Systems Technology and Design Pillar, Singapore University of Technology and Design during 2018-2019. His research interests include formal methods, software engineering, cyber-security and machine learning.
 \end{IEEEbiography}

 \begin{IEEEbiography}[{\includegraphics[width=1in,height=1.25in,clip,keepaspectratio]{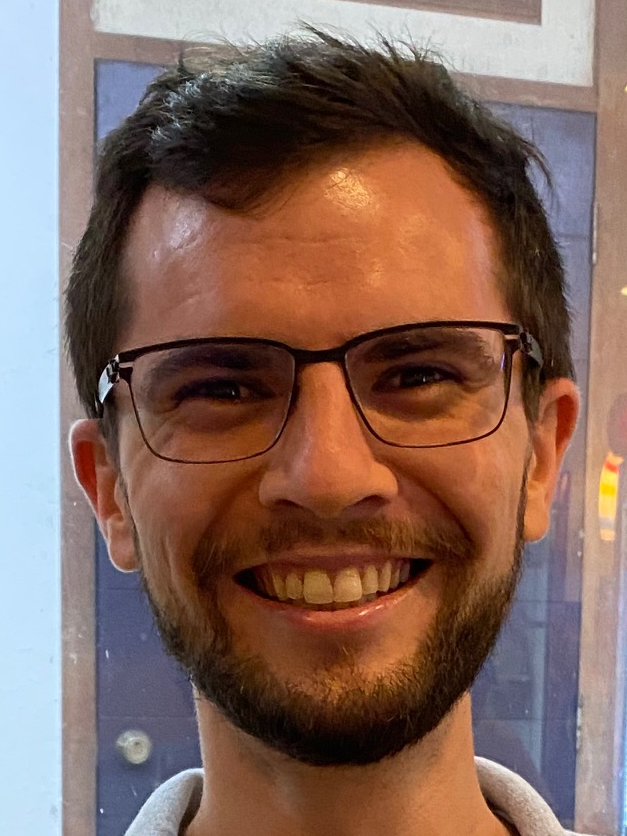}}]{Christopher M. Poskitt} is an Associate Professor of Computer Science (Education) at Singapore Management University (SMU), where he is part of the Centre for Research on Intelligent Software Engineering. Prior to SMU, he held postdoctoral research positions at ETH Z\"{u}rich and SUTD, and obtained his PhD in Computer Science from the University of York (2014). His research broadly addresses the problem of engineering correct and secure software, especially in the context of cyber-physical systems (e.g.~industrial control systems, autonomous vehicles). In addition to software engineering, his research interests span formal methods, cybersecurity, and computer science education.
 \end{IEEEbiography}

 \begin{IEEEbiography}[{\includegraphics[width=1in,height=1.25in,clip,keepaspectratio]{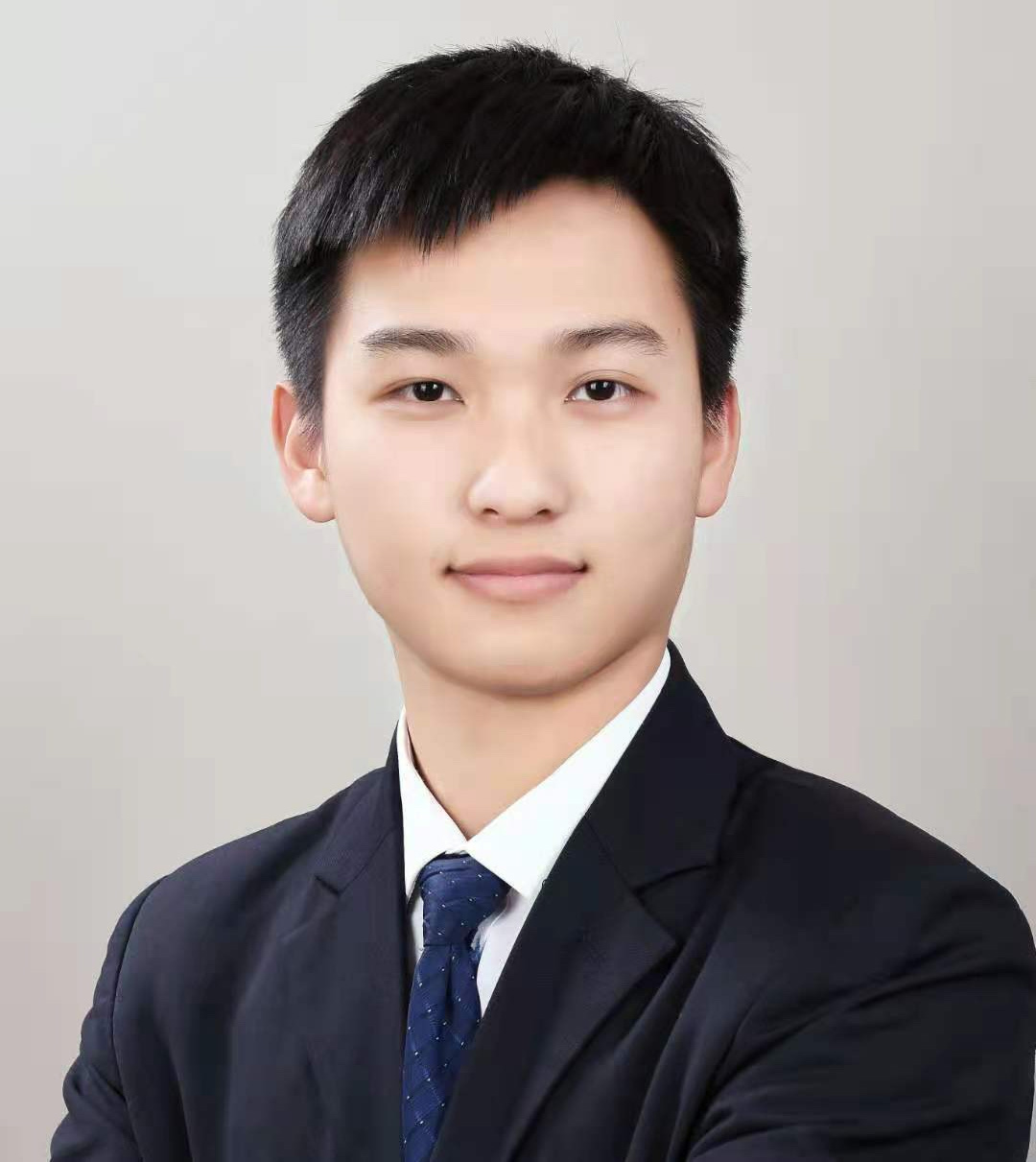}}]{Xiangxiang Chen} received the B.Eng. degree in mechanical engineering from Xi'an Jiaotong University, Xi'an, China in 2021. He is working toward the Ph.D degree in Cyberspace Security at the IS2 Lab at School of Control Science and Engineering, Zhejiang University, Hangzhou, China. His research interests include fuzzing and AI system testing.
 \end{IEEEbiography}

 \begin{IEEEbiography}[{\includegraphics[width=1in,height=1.25in,clip,keepaspectratio]{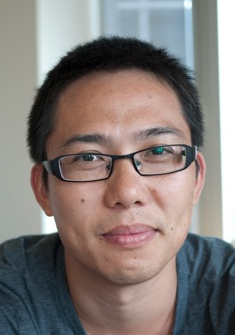}}]{Jun Sun} is currently a tenured professor at the School of Information Systems, Singapore Management University. He received bachelor’s and Ph.D. degrees in computing science from the National University of Singapore (NUS) in 2002 and 2006, respectively. From 2010 to 2019, he was an assistant/associate professor at the Singapore University of Technology and Design. He was a visiting scholar at MIT from 2011 to 2012. His research focuses on software engineering, formal methods, program analysis, and cyber-security. He is the co-founder of the PAT model checker.
 \end{IEEEbiography}

 \begin{IEEEbiography}[{\includegraphics[width=1in,height=1.25in,clip,keepaspectratio]{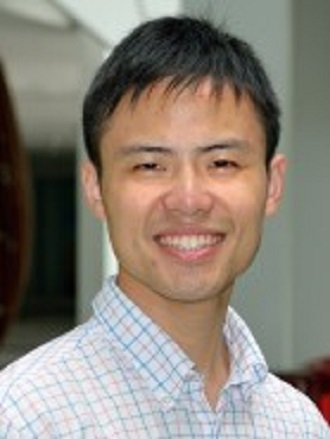}}]{Peng Cheng} received the B.Sc. degree in automation and the Ph.D. degree in control science and engineering, from Zhejiang University, Hang Zhou, China, in 2004 and 2009, respectively. From 2012 to 2013, he worked as Research Fellow in Information System Technology and Design Pillar, Singapore University of Technology and Design. He is currently a Professor with the College of Control Science and Engineering, Zhejiang University, Hangzhou, China. His research interests include networked sensing and control, cyber-physical systems, and control system security.
 \end{IEEEbiography}







\end{document}


%

\appendices
\section{Inconsistency Caused by $\mathtt{RETURN;}$ and $\mathtt{EXIT;}$}
\label{ap1}
In this section, we show case studies of inconsistent running results caused by ``$\mathtt{RETURN;}$'' and ``$\mathtt{EXIT;}$''. In Code 1, function block $\mathtt{ADD}$ performs a simple summation operation. It involves three variables $\mathtt{N}$, $\mathtt{M}$ and $\mathtt{Sum}$. The program adds the modified $\mathtt{N}$ and $\mathtt{M}$ to $\mathtt{Sum}$ on line 10. It is clear that this is an illegal operation because it assigns integer values to variables of $\mathtt{REAL}$ types. However, due to the $\mathtt{RETURN;}$ instruction on the 9th line, in fact, function block $\mathtt{ADD}$ will never execute to the 10th line, and will return the calling POU on the 9th line. This makes the executable semantics unable to detect the exception that occurs on line 10. A similar situation will be generated by the $\mathtt{EXIT;}$, as shown in code 2.

\begin{lstlisting}[caption = Example of $\mathtt{RETURN}$,numbers=left]
FUNCTION_BLOCK ADD
    VAR
        N : INT := 100;
        M : INT := 35;
        Sum : REAL := 0;
    END_VAR
    
    N := N - (N / 2);
    RETURN;
    Sum := N + M;
END_FUNCTION_BLOCK
\end{lstlisting}

\begin{lstlisting}[caption = Example of $\mathtt{EXIT}$,numbers=left]
FUNCTION_BLOCK ADD
    VAR
        N : INT := 0;
        Sum : REAL := 0;
    END_VAR
    
    WHILE N < 10 DO
        Sum := Sum + N;
        EXIT;
        N := N + 2.0;
    END_WHILE;
END_FUNCTION_BLOCK
\end{lstlisting}

\section{Bugs of OpenPLC}
\label{ap2}
In this section, we show case studies of bugs which showed in TABLE 8 in OpenPLC.
\subsection{``VAR'' Parsing Exception}
Code 3 shows a very simple program $\mathtt{Declaration}$, which declares an integer variable named $\mathtt{var0}$ and then assigns it to 3. The correctness of the program is actually obvious, but OpenPLC will report an error on line 6 when compiling.

\begin{lstlisting}[caption = ``VAR'' parsing exception]
PROGRAM Declaration
    VAR
        var0 : INT := 0;
    END_VAR
    
    var0 := 3;
END_PROGRAM
\end{lstlisting}

\subsection{Division by Zero}
Code 4 shows a simple division operation. Program $\mathtt{Division}$ contains two variables, $\mathtt{N}$ and $\mathtt{M}$. OpenPLC can detect and report division by zero when executing line 7, but compile and run normally when executing line 8. In fact, the value of $\mathtt{M}$ is also 0.

\begin{lstlisting}[caption = Division by zero]
PROGRAM Division
    VAR
        N : INT := 5;
        M : INT := 0;
    END_VAR
    
    //N := N / 0;
    N := N / M;
END_PROGRAM
\end{lstlisting}

\subsection{Overflow Access}
Program $\mathtt{Array\_access}$ in code 5 contains a declaration of a list and access to the elements in the array. When we use the instructions on lines 8 and 10 for member access, OpenPLC can detect out-of-bounds. But when we use lines 9 and 11, OpenPLC supports the program to execution. However, they are essentially the same.

\begin{lstlisting}[caption = Overflow access]
PROGRAM Array_access
    VAR
        N : ARRAY [1..3] OF INT := [2,3,4];
        Min : INT := 0;
        Max : INT := 6;
    END_VAR
    
    //Min := N[0];
    Min := N[Min];
    //Max := N[6];
    Max := N[Max];
END_PROGRAM
\end{lstlisting}

\subsection{MOD by Zero and MOD Exception}
Program $\mathtt{Mod\_Exception}$ in Code 6 contains some simple modulation operations. For the $5 \mathtt{MOD} 0$ in line 6, there is no reasonable operation result, but OpenPLC provides this calculation and the result is 0. For the $\ \mathtt{MOD} (165)$ in line 7, it is obviously wrong in theory, but OpenPLC can also calculate and the result is 0.

\begin{lstlisting}[caption = MOD by zero and MOD Exception]
PROGRAM Mod_Exception
    VAR
        N : INT := 0;
    END_VAR
    
    N := 5 MOD 0;
    N := MOD (165);
END_PROGRAM
\end{lstlisting}